\documentclass[%
 superscriptaddress,
 twocolumn,
 nofootinbib,
 amsmath,amssymb,
 aps,
 prd,
]{revtex4-2}

\usepackage[colorlinks]{hyperref}
\usepackage{tabularx}
\usepackage{graphicx}
\usepackage{amssymb,amsmath,bm,tensor,braket}
\usepackage{xcolor}
\usepackage[varg]{txfonts}
\usepackage{enumerate}
\usepackage{mathtools}
\usepackage[capitalize]{cleveref}
\usepackage[utf8]{inputenc}
\usepackage[normalem]{ulem}
\usepackage{dcolumn}

\usepackage{graphicx}	
\usepackage{amsmath}	

\newcommand{\Reply}[1]{#1}

\newcommand{\s}{\mathrm s}
\newcommand{\n}{\mathrm n}
\newcommand{\p}{\mathrm p}
\newcommand{\e}{\mathrm e}
\newcommand{\dd}{\mathrm d}

\usepackage{subfigure}
\usepackage[normalem]{ulem}

\def\prd{Phys.~Rev.~D}       
\def\prl{Phys.~Rev.~Lett}    
\def\aap{A\&A}                

\begin{document}

\title{$g$-mode of neutron stars in pseudo-Newtonian gravity}

\author{Hong-Bo Li}
\affiliation{Department of Astronomy, School of Physics, Peking University, Beijing 100871, China}%
\affiliation{Kavli Institute for Astronomy and Astrophysics, Peking University, Beijing 100871, China}

\author{Yong Gao}%
\affiliation{Department of Astronomy, School of Physics, Peking University, Beijing 100871, China}%
\affiliation{Kavli Institute for Astronomy and Astrophysics, Peking University, Beijing 100871, China}

\author{Lijing Shao}\email[Corresponding author: ]{lshao@pku.edu.cn}
\affiliation{Kavli Institute for Astronomy and Astrophysics, Peking University, Beijing 100871, China}%
\affiliation{National Astronomical Observatories, Chinese Academy of Sciences, Beijing 100012, China}

\author{Ren-Xin Xu}\email[Corresponding author: ]{r.x.xu@pku.edu.cn}
\affiliation{Department of Astronomy, School of Physics, Peking University, Beijing 100871, China}%
\affiliation{Kavli Institute for Astronomy and Astrophysics, Peking University, Beijing 100871, China}%

\date{\today}

\begin{abstract}
The equation of state (EOS) of nuclear dense matter plays a crucial role in many
astrophysical phenomena associated with neutron stars (NSs).  Fluid oscillations
are one of the most fundamental properties therein.  NSs support a family of
gravity $g$-modes, which are related to buoyancy.  We study the gravity
$g$-modes caused by composition gradient and density discontinuity in the
framework of pseudo-Newtonian gravity. The mode frequencies are calculated in
detail and compared with Newtonian and general-relativistic (GR) solutions.  We
find that the $g$-mode frequencies in one of the pseudo-Newtonian treatments can
approximate remarkably well the GR solutions, with relative errors in the order
of $1\%$.  Our findings suggest that, with much less computational cost,
pseudo-Newtonian gravity can be utilized to accurately analyze oscillation of
NSs constructed from an EOS with a first-order phase transition between nuclear
and quark matter, as well as to provide an excellent approximation of GR effects
in core-collapse supernova (CCSN) simulations. 
\end{abstract}

\maketitle

\section{Introduction}\label{sec:Introduction}

The oscillation modes of neutron stars (NSs) provide a means to probe the
internal composition and state of dense matter.  NSs have rich oscillation
spectra, with modes associated with different physical origins, such as
the internal ingredients, the elasticity of the crust, superfluid components, 
and so on \cite{Andersson:2019}.  For typical non-rotating fluid stars, the
oscillation modes include the fundamental ($f$), pressure ($p$), and gravity
($g$) modes, which provided the basic classification of modes according to the
physics dominating their behaviours \cite{Cowling:1941}. More realistic stellar
models and rotation introduce additional classes of oscillation modes.

In this work, we study the $g$-mode oscillations for non-rotating NSs in the
framework of pseudo-Newtonian gravity \cite{Marek:2005if, Mueller:2008it,
Yakunin:2015wra, Morozova:2018glm, OConnor:2018sti, OConnor:2015rwy,
Zha:2020gjw, Tang:2021woo}.  \citet{Reisenegger:1992} investigated the $g$-mode
induced by composition (proton-to-neutron ratio) gradient in the cores of NSs.
Moreover, hot young NSs may excite $g$-modes supported by entropy gradients
\cite{McDermott:1983, McDermott:1988, Ferrari:2003nk, Kruger:2014pva}.  
It has
also been demonstrated that the onset of superfluidity has a key influence on
the buoyancy that supports the $g$-modes \cite{Lee:1995, Gusakov:2013eoa, Kantor:2014lja,
Andersson:2001bz, Passamonti:2015oia}.  Density discontinuity produced by abrupt composition
transitions may play an important role in determining the $g$-mode properties
\cite{Finn:1987, McDermott:1990}.  \citet{Sotani:2001bb} calculated $f$ and $g$
modes of NSs with density discontinuity at an extremely high density and
discussed the stability of the stellar models.  A phase transition occurred in
the cores of NSs with a polytropic equation of state (EOS) has been studied by
\citet{Miniutti:2002bh}.  The frequencies of $g$-modes from density
discontinuity are larger than those induced by the entropy gradient.  Furthermore,
discontinuity $g$-mode may occur in perturbed quark-hadron hybrid stars
\cite{Tonetto:2020bie, Constantinou:2021hba}.  Recently, \citet{Zhao:2022toc}
considered the $g$-mode of NSs containing quark matter and discussed the Cowling
approximation, which leads to a relative error of $\sim 10\%$ for higher-mass
hybrid stars.  We here focus on the $f$ and $g$ modes of NSs in pseudo-Newtonian
gravity caused by the first-order phase transition in the cores of NSs.

The study of NS oscillations is timely in the gravitational-wave era 
\cite{LIGOScientific:2017vwq, LIGOScientific:2018cki, Li:2022qql}.  Tidal
interaction in a coalescing binary NS can resonantly excite the $g$-mode
oscillation of NSs when the frequency of the tidal driving force approaches the
$g$-mode frequencies \cite{Lai:1993di, Kuan:2021jmk}.  Moreover, the mixture of
pure-inertial and inertial-gravity modes can become resonantly excited by
tidal fields for rotating NSs \cite{Lai:2006pr, Xu:2017hqo}.  The $g$-mode can
also result in secular instability in rotating NSs \cite{Lai:1998yc}.
\citet{Gaertig:2009rr} considered the $g$-mode of fast-rotating stratified NSs
using the relativistic Cowling approximation. The typical scenarios pertain to
the $p$-$g$ mode instability and the saturation of unstable modes
\cite{Weinberg:2013pbi, LIGOScientific:2018ehx}.  The universal relation of
$g$-mode asteroseismology has been discussed by \citet{Kuan:2022bhu} for
different classes of EOSs. In particular, the absence of very low-frequency
$g$-modes helps to explain the absence of tidal resonances
\cite{Andersson:2019mxp}.  The cut-off in the high-order $g$-mode spectrum may
also be relevant for  scenarios of nonlinear mode coupling.  The properties of
$g$-modes for newly-born strange quark stars and NSs using Cowling approximation
in Newtonian gravity have been discussed by \citet{Fu:2008bu}.

Hydrodynamical simulations are necessary to study the properties of the proto-NS
in a core-collapse supernova (CCSN). The $g$-mode of such a scenario may impact
associated gravitational waves~\cite{Ott:2006qp}. However, the physics of
neutrino transport and EOS is very uncertain for the hydrodynamical simulations.
As multi-dimensional general-relativistic (GR) codes for numerical simulations
are scarce and have high demand of computational cost, most previous
investigations relied on the Newtonian approximation for the strong
gravitational field and fluid dynamics \cite{Marek:2005if, Mueller:2008it}.
Nevertheless, ``Case A potential'' formalism (c.f.\ Sec.~\ref{sec: Case A potentia}) was found to 
be  a good approximation to relativistic solutions in
simulating non-rotating or slowly rotating CCSNs.  This potential allows for an
accurate approximation of GR effects in an otherwise Newtonian hydrodynamic
code, and it also works for cases of rapid rotation \cite{Mueller:2008it}.
This has motivated a sequence of CCSN simulations \cite{Yakunin:2015wra,
Morozova:2018glm, OConnor:2018sti, OConnor:2015rwy}. 
The effectiveness of using Case A potential formalism to approximate GR has been studied by
\citet{Mueller:2008it, Pajkos:2019nef, OConnor:2018sti}. 
In particular, \citet{Mueller:2008it} found that 
Case A potential formalism can not obtain the correct oscillation modes and 
indicated the failure of the Case A potential, possibly being attributed to the absence of a lapse function.
Recently, \citet{Zha:2020gjw} have extended the Case A potential formalism 
with a lapse function to simulate the oscillation of proto-neutron star (PNS).  
They found that Case A potential formalism with an additional 
lapse function can approximate well the frequency of the fundamental radial mode.

\citet{Tang:2021woo} studied the 
radial and non-radial oscillation modes of NSs in pseudo-Newtonian gravity,
including the Case A potential with and without the lapse function.  Motivated
by \citet{Tang:2021woo}, we here study the $g$-mode of NS cores using Case A
potential formalism with and without the lapse function. Our findings suggest
that, with much less computational cost, pseudo-Newtonian gravity can be
utilized to accurately analyze oscillation of NSs constructed from an EOS with a
first-order phase transition, thus to provide an excellent approximation
of GR effects in CCSN simulations. 

The paper is organized as follows. In Sec.~\ref{sec: Key_ingredient}, we
introduce the key ingredients of the model, including different pseudo-Newtonian
schemes and the buoyancy nature associated with $g$-mode.  The local dynamics of
NS cores, including composition gradient and density discontinuity, are
presented in Sec.~\ref{sec: NR}.  Finally, we summarize our work in
Sec.~\ref{sec: conculsion}.  Throughout the paper, we adopt geometric units with
$c=G=1$, where $c$ and $G$ are the speed of light and the gravitational
constant, respectively.

\section{Key ingredients of the model}\label{sec: Key_ingredient}

\subsection{Case A potential in pseudo-Newtonian gravity}\label{sec: Case A potentia}

Case A effective potential is defined by replacing the Newtonian gravitational
potential in a spherically symmetric Newtonian hydrodynamic simulation by
\cite{Marek:2005if, Tang:2021woo}
\begin{equation} \label{eq:phiTOV}
    \Phi_{\rm TOV}(r)=-4\pi\int^\infty_r \frac{\dd r^{\prime}}{r^{\prime 2}} \left(\frac{m_{\rm TOV}}{4\pi}+r^{\prime 3} P\right) \times \frac{1}{\Gamma^2} \left(\frac{\rho+\rho\varrho+P}{\rho}\right)\,,
\end{equation}
where $r$ is the radial coordinate, $\rho$ is the rest-mass density, $P$ is the
pressure, $\varrho$ is the specific internal energy, and the total energy
density is given by $\epsilon=\rho+\rho\varrho$.  The function $m_\text{TOV}$ is
defined by
\begin{equation} \label{eq:mTOV}
    m_{\rm TOV}(r) = 4\pi\int^r_0 \dd r^{\prime} r^{\prime 2} \epsilon \Gamma \,,
\end{equation}
with
\begin{equation}
    \Gamma=\sqrt{1-2\frac{m_{\rm TOV}}{r}} \,.
    \label{eq: Gamma}
 \end{equation}
From Eq.~(\ref{eq:phiTOV}) and Eq.~(\ref{eq:mTOV}), we have  
\begin{align}
\frac{{\rm d} m_{\rm TOV}}{{\rm d} r} &= 4\pi r^2\epsilon\Gamma 
\label{eq: def_1}\,, \\
\frac{{\rm d} \Phi_{\rm TOV}}{{\rm d} r} &= \frac{4\pi}{r^2}\left(\frac{m_{\rm TOV}}{4\pi}+r^3 P\right)
\frac{1}{\Gamma^2}\frac{(\epsilon+P)}{\rho} 
\label{eq: def_2}\,.
\end{align}
We use the Case A and Case A+lapse schemes and the other four schemes to study
the $g$-mode originating from the composition gradient and density discontinuity
of NS cores in the framework of pseudo-Newtonian gravity.  All background and
perturbation equations for each scheme are given in the next three subsections
and summarized in Table \ref{tab: schemes}.

\begin{table}
	\centering
	\caption{Different schemes to calculate the oscillation modes, along with
	the corresponding background and the lapse function. Non-radial perturbation
	equations are the same [Eqs.~(\ref{eq: bc_p1}--\ref{eq: bc_p4})] for all six
	schemes, but some of them include a lapse-function $\alpha$ in the
	hydrodynamic equations.  Note that the lapse function only appears in the
	perturbation equations but not in the background equations.}
    \renewcommand\arraystretch{1.3}
	\begin{tabular}{llc}
		\hline
		Scheme & Background equations &  Lapse function $\alpha$ \\
		\hline
		N & Eqs. (\ref{eq: N_dmdr}) to (\ref{eq: N_dphidr}) & --\\
		N+lapse & Eqs. (\ref{eq: N_dmdr}) to (\ref{eq: N_dphidr})  & Eq. (\ref{eq: lapse})\\
		TOV & Eqs. (\ref{eq: GR_dmdr}) to (\ref{eq: GR_dphidr}) & --\\
		TOV+lapse & Eqs. (\ref{eq: GR_dmdr}) to (\ref{eq: GR_dphidr}) & Eq. (\ref{eq: lapse})\\
		Case A & Eqs. (\ref{eq: A_dmdr}) to (\ref{eq: A_dphidr}) & --\\
		Case A+lapse & Eqs. (\ref{eq: A_dmdr}) to (\ref{eq: A_dphidr})  & Eq. (\ref{eq: lapse})\\
		\hline
	\end{tabular}
	\label{tab: schemes}
\end{table}

\subsection{Equilibrium configurations}\label{sec: EC}

We consider the following three sets of equilibrium configurations.
\begin{enumerate}[(I)]
  \item For the Newtonian (N) and Newtonian+lapse function (N+lapse) schemes, 
  the hydrostatic equilibrium equations are
    \begin{align}
    \frac{{\rm d} m}{{\rm d} r} &= 4\pi r^2\rho
    \label{eq: N_dmdr} \,, \\
    \frac{{\rm d} P}{{\rm d} r} &= -\frac{\rho m}{r^2} 
    \label{eq: N_dpdr} \,, \\
    \frac{{\rm d} \Phi}{{\rm d} r} &= -\frac{1}{\rho}\frac{{\rm d} P}{{\rm d} r}
    \label{eq: N_dphidr} \,.
    \end{align}
    where $\rho$ is the rest-mass density, and $\Phi$ is the Newtonian gravitational
    potential.
  \item Instead, if we consider spherical and static stars in GR, we have the
  Tolman-Oppenheimer-Volkoff (TOV) equations
    \begin{align}
    \frac{{\rm d} m}{{\rm d} r} &= 4\pi r^2\epsilon
    \label{eq: GR_dmdr} \,, \\
    \frac{{\rm d} P}{{\rm d} r} &= -\frac{(\epsilon+P)(m+4\pi r^3 P)}{r(r-2m)}
    \label{eq: GR_dpdr} \,, \\
    \frac{{\rm d} \Phi}{{\rm d} r} &= -\frac{1}{\epsilon+P}\frac{{\rm d} P}{{\rm d} r}
    \label{eq: GR_dphidr} \,.
    \end{align}
  \item Lastly, for the Case A and Case A+lapse schemes, the background
  equations are obtained by replacing the Newtonian gravitational potential by
  the Case A potential \cite{Marek:2005if, Tang:2021woo}, and we have
\begin{align}
\frac{{\rm d} m}{{\rm d} r} &= 4\pi r^2\epsilon\Gamma
\label{eq: A_dmdr} \,, \\
\frac{{\rm d} P}{{\rm d} r} &= -\frac{4\pi}{r^2}\left(\frac{m}{4\pi}+r^3 P\right)\frac{1}{\Gamma^2}(\epsilon+P)
\label{eq: A_dpdr} \,, \\
\frac{{\rm d} \Phi}{{\rm d} r} &= -\frac{1}{\rho}\frac{\dd P}{\dd r} 
\label{eq: A_dphidr} \,.
\end{align}
\end{enumerate}
We use the modified Newtonian hydrodynamic equations in \citet{Zha:2020gjw} and \citet{Tang:2021woo},
where a lapse function $\alpha$ is added to mimic the time-dilation effect. 
The modified hydrodynamic equations are:
\begin{align}
	&\frac{\partial \rho}{\partial t}+\nabla\cdot(\alpha \rho \vec{v})=0 , \label{Euler1_mod}\\
	&\frac{\partial}{\partial t}(\rho\vec{v})+\nabla\cdot[\alpha(\rho\vec{v}\vec{v}+P \overset\leftrightarrow{I})]=-\alpha(\rho-P)\nabla\Phi \,, \label{Euler2_mod}
\end{align}
where $\vec{v}$ is the fluid velocity and the lapse function
is defined by
\begin{equation}\label{eq: lapse}
 \alpha=\exp (\Phi) \,.
\end{equation}
The readers can infer \citet{Tang:2021woo} for a detailed variational derivation of the linearized fluid equations. We will use the same lapse function in our calculations.

\subsection{Buoyancy and the $g$-mode}\label{sec: g-mode}

As well known that NSs always have real frequency $f$-mode and $p$-mode regimes.
However, $g$-mode may have a real, imaginary,  and zero frequency, which
correspond to convective stability, instability, and marginal stability.  We
consider the local dynamics of NS cores, focusing on the buoyancy experienced by
fluid elements and the associated $g$-mode.  The frequencies of $g$-modes are
closely related to the Brunt-V\"ais\"al\"a frequency $N$, defined via
\begin{equation}\label{eq: buoyancy}
  N^2=g_{N}^2\left(\frac{1}{c_\e^2}-\frac{1}{c_\s^2}\right) \,,
\end{equation}
where $g_{N}$ is the positive Newtonian gravitational acceleration, $c_\s$
is the adiabatic sound speed, 
\begin{equation}\label{eq: adiabatic sound speed}
  c_\s^2= \left(\frac{\partial P}{\partial \rho}\right)_{\s}  \,.
\end{equation}
Here the subscript ``s'' means ``adiabatic'', which in this case implies constant composition.
The quantity $c_\e$ is given by 
\begin{equation}\label{eq: equilibrium sound speed}
  c_\e^2= \frac{{\rm d} P}{{\rm d} \rho}  \,,
\end{equation}
where the subscript ``e'' stands for ``equilibrium''.
If $c_\s^2 = c_\e^2$, the star exhibits no convective phenomena (zero-buoyancy
case).  In this work, we consider only the $g$-mode of NS cores, so we set
$c_\s^2 = c_\e^2$ for the crustal region.  Again, $c_\s^2 > c_\e^2$ ($c_\s^2 <
c_\e^2$) denotes convective stability (instability).  Combining Eqs.~(\ref{eq:
buoyancy}--\ref{eq: equilibrium sound speed}), we can write the
Brunt-V\"ais\"al\"a frequency as
\begin{equation}\label{eq: bf}
 N^{2}= -A g_{N} \,,
\end{equation}
where $A$ is 
\begin{equation}\label{eq: S_discriminant_1}
 A= \frac{{\rm d} \ln\rho}{{\rm d} r}-  \frac{1}{\Gamma_{1}} \frac{{\rm d} \ln P}{{\rm d} r} \,,
\end{equation}
which is called the Schwarzschild discriminant.  If the star model obeys a
simple polytropic EOS, $P=K\rho^{\gamma}$, then $\gamma=\rm d\ln P/ \rm d\ln\rho$
is defined for the unperturbed background configuration.  Hence, the
Schwarzschild discriminant becomes
\begin{equation}\label{eq: S_discriminant_2}
 A= \left(\frac{1}{\gamma} - \frac{1}{\Gamma_{1}}\right) \frac{{\rm d} \ln P}{{\rm d} r} \,.
\end{equation}
Clearly, if the adiabatic index $\Gamma_{1}>\gamma$, the star is related to the
convective stability, in the case of $c_\s^2 > c_\e^2$.  In Sec.~\ref{sec:
Composition gradient}, we will calculate the frequencies of $g$-modes for the
composition gradient, which is related to the discussion here.

\subsection{Non-radial perturbation equations}\label{sec: nonradial_osc}

In this section, we study non-radial oscillations of NSs in pseudo-Newtonian
gravity. \citet{Tang:2021woo} calculated the quadrupole ($\ell=2$) $f$ and $p$
modes.  The perturbation of scalars is expanded in spherical harmonics and the
Lagrangian displacement is expanded in vector spherical harmonics
\cite{McDermott:1988, Tang:2021woo}. When considering an eigenmode, we have
\begin{align}
&\delta\rho  = \delta\tilde{\rho}(r)Y_{\ell m}\,,\\
&\delta P  = \delta\tilde{P}(r)Y_{\ell m} \,, \\
&\delta\Phi =  \delta\tilde{\Phi}(r)Y_{\ell m} \,, \\
&\vec{\xi}  = U(r) Y_{\ell m} {\hat r}  +V(r) \nabla Y_{\ell m} \,,
\end{align}
where $Y_{\ell m}$ is the standard spherical harmonic function, and $\hat r$ is
the radial unit vector.  Then one can obtain the following system of equations
for the fluid perturbations (see \citet{Tang:2021woo}, for a detailed variational derivation),

\begin{align}
\frac{{\rm d} U}{{\rm d} r}&=-\left(\frac{2}{r}+\frac{{\rm d}\Phi}{{\rm d}r}+\frac{1}{\gamma P}\frac{{\rm d}P}{{\rm d}r} - \frac{A}{\alpha}\right)U 
        + \left[\frac{\alpha \ell(\ell+1)}{\rho r^2 \omega^2}  -  \frac{1}{\alpha\Gamma_1 P}\right]\delta\tilde{P}  \nonumber\\ 
    & \quad + \frac{\alpha \ell(\ell+1)}{r^2 \omega^2}\delta\tilde{\Phi}  \,,
\label{eq: bc_p1} \\
\frac{{\rm d}\delta\tilde{P}}{{\rm d} r}&=\left(\frac{\rho\omega^2}{\alpha} - \frac{{\rm d} P}{{\rm d} r}A\right)U + \frac{1}{\Gamma_1 P}\frac{{\rm d} P}{{\rm d} r}\delta\tilde{P} - \rho\frac{{\rm d} \delta\tilde{\Phi}}{{\rm d}r}  \,,  \label{eq: bc_p2} \\
\frac{{\rm d} \delta\tilde{\Phi}}{{\rm d} r}&=\Psi  \,,
\label{eq: bc_p3}\\
\frac{{\rm d} \Psi}{{\rm d} r}&=-\frac{2}{r}\Psi + \frac{\ell(\ell+1)}{r^2}\delta\tilde{\Phi} + 4\pi\frac{\rho}{\Gamma_1 P}\delta\tilde{P} - 4\pi\rho AU   \,.
 \label{eq: bc_p4} 
\end{align}

To solve these equations, we require the boundary conditions at the center and
surface of the NS.  At the center,  the regularity conditions of the variables
yield the following relations \cite{Westernacher-Schneider:2020bkw, Tang:2021woo}
\begin{align}
    & U = r^{\ell-1}A_{0}  \,, \\
    & \delta\tilde{P} = r^{\ell} B_{0} \,, \\
    & \delta\tilde{\Phi} = r^{\ell} C_{0}  \,, \\
    & \Psi = \ell r^{\ell-1} C_{0} \,, \\
    & A_{0} = \frac{\alpha \ell}{\rho\omega^{2}} (B_{0} + \rho C_{0}) \,,
\end{align}
where $B_{0}$ and $C_{0}$ are constants.  At the surface of the star, the
perturbed pressure must vanish, which provides 
\begin{equation}
  \frac{{\rm d} P}{{\rm d} r}U+\delta\tilde{P}=0  \label{eq: surface_BC1} \,.
\end{equation}
The $\delta\tilde{\Phi}$ and ${{\rm d}\delta\tilde{\Phi}}/{{\rm d}r}$ are
continuous, so we obtain
\begin{equation}
  \Psi=-\frac{\ell+1}{r}\delta\tilde{\Phi}  \label{eq: surface_BC2}  \,.
\end{equation}
Note that in the N, Case A, and TOV schemes, the lapse function equals to 1
($\alpha=1$).

\begin{table*}
    \centering
    \caption{Comparison of the non-radial mode frequencies (unit: Hz) of a
    polytropic star model where polytropic index $\gamma=2$, $K=1.4553 \times
    10^{5} \ \rm g^{-1}\,cm^{5}\,s^{-2}$, and central density
    $\rho_{c}=7.9\times10^{14}\ \rm g\,cm^{-3}$, to earlier results of \citet{Westernacher-Schneider:2020bkw} and \citet{Tang:2021woo}.}
        \renewcommand\arraystretch{1.3}
        \begin{tabular}{c c c c c c c }
        \hline
        Mode &  \citet{Westernacher-Schneider:2020bkw} & \citet{Tang:2021woo} & $\Gamma_{1}=2.01$ &
        $\Gamma_{1}=2.05$ &$\Gamma_{1}=2.1$ & $\Gamma_{1}=2.15 $ \\
        \hline
        $p_{2}$ & 7290 & 7932 & 7957 & 8049 & 8163 & 8276 \\
        $p_{1}$ & 5122 & 5131 & 5151 & 5216 & 5297 & 5377 \\
        $f$         & 2024 & 2021 & 2021 & 2025 & 2029 & 2032 \\
        $g_{1}$ & --   & --   & 143  & 317 & 441 & 532 \\
        $g_{2}$ & --   & --   & 99   & 219 & 306 & 369 \\
        $g_{3}$ & --   & --   & 76   & 169 & 235 & 284\\
        \hline
    \end{tabular}
    \label{tab: Tang_compare}
\end{table*}

 To test our numerical code, we have redone calculations with the same
 polytropic EOS as that in the Appendix A of \citet{Marek:2005if}, where the
 polytropic index $\gamma$ and the adiabatic index $\Gamma_{1}> \gamma$ are
 constant throughout the stellar interior.  Detailed numerical results are shown
 in Table \ref{tab: Tang_compare}.  It is noted that our numerical results for
 the polytropic model with $\Gamma_{1}= \gamma$ agree with Table 3 of
 \citet{Tang:2021woo}.  In Table \ref{tab: Tang_compare}, we compare the
 frequencies of $p$, $f$, and $g$ modes computed with $\Gamma_{1}> \gamma$ and
 $\Gamma_{1}= \gamma$ \cite{Westernacher-Schneider:2020bkw, Tang:2021woo}.  The
 frequencies of $p$ and $f$ modes increase with the increase of the adiabatic
 index $\Gamma_{1}$.  In particular, the $g$-mode frequencies also increase with
 increase of the adiabatic index $\Gamma_{1}$, which indicates a larger buoyancy.

\section{NUMERICAL RESULTS}\label{sec: NR}

\subsection{Composition gradient}\label{sec: Composition gradient}

Taking the matter composition into account, and assuming that the model accounts
for the presence of neutrons, protons, and electrons, we have a two-parameter
EOS, $P=P(n,x)$, which is a function of the baryon number density $n$ and the
proton fraction $x=n_\p/n$.  Specifically, we use shorthand notations: ``$\n$'' for
neutrons, ``$\p$'' for protons, and ``$\e$'' for electrons.  The energy per baryon of the
nuclear matter can be written as \cite{Lagaris:1981, Prakash:1988,
Wiringa:1988tp, Lai:1993di}
\begin{equation} \label{eq:Ennx}
   E_n(n,x)=T_n(n,x)+V_{0}(n)+V_{2}(n)(1-2x)^2 \,,
\end{equation}
where
\begin{equation}
   T_n(n,x)={\frac{3}{5}\frac{\hbar^2}{2m_\n} (3\pi^2n)^{2/3}[x^{5/3}}+(1-x)^{5/3}] \,,
\end{equation}
is the Fermi kinetic energy of the nucleons, and $m_n$ is the nucleon mass.
$V_{0}$ mainly specifies the bulk compressibility of the matter, and $V_{2}$ is
related to the symmetry energy of nuclear matter \cite{Lattimer:2014scr}. 

To compare the results of $g$-modes in Newtonian gravity \cite{Lai:1993di}, we
adopt the same $V_{0}$ and $V_{2}$ for different EOS models, based on the
microscopic calculations in \citet{Wiringa:1988tp}.  Detailed numerical results
of $V_{0}$ and $V_{2}$ have been tabulated in Table IV of
\citet{Wiringa:1988tp}.  The approximate formulae of $V_{0}$ and $V_{2}$ are
presented in Sec.~4.3 of \citet{Lai:1993di}.

In this work, we consider the model ``AU'' 
(the EOS based on nuclear potential AV14+UVII in \citet{Wiringa:1988tp})
and the model ``UU'' 
(the EOS based on nuclear potential UV14+UVII in \citet{Wiringa:1988tp}), respectively. 
For the model AU,  $V_0$ and $V_2$ (in the unit of MeV) are fitted as \cite{Lai:1993di}
\begin{align}
V_{0}& = -43+330\,(n-0.34)^2  \label{eq: AU_1} \,, \\
V_{2}& = 21\,n^{0.25} \label{eq: AU_2} \,,
\end{align}
where $n$ is the baryon number density in $\rm fm^{-3}$.  For the model UU, we
have 
\begin{align}
V_{0}& = -40+400\,(n-0.3)^2  \label{eq: UU_1} \,, \\
V_{2}& = 42\,n^{0.55} \label{eq: UU_2} \,.
\end{align}
These fitting formulae are valid for $0.07 \, {\rm fm}^{-3} \leq n \leq 1\, \rm fm^{-3} $.  
For densities $0.001 \, {\rm fm}^{-3} < n <0.07 \, \rm fm^{-3}$, we employ the EOS of \citet{BBP:1971},  
while for $n \leq 0.001 \, \rm fm^{-3}$, we employ the EOS of \citet{BPS:1971}.

Once we have this relation, we can work out the mass-energy density, pressure, 
and adiabatic sound speed. The equilibrium configuration must satisfy the beta equilibrium,
\begin{equation}
  \mu_\n = \mu_\p+\mu_\e \,, \label{eq: one}
\end{equation}
and the charge neutrality
\begin{equation}
n_\p = n_\e \,,  \label{eq: two}
\end{equation}
where $\mu_{i} $ are the chemical potentials of the three species of particles. 
The equilibrium proton fraction $x(n) = x_{\e}(n)$ can be obtained by solving
Eqs.~(4.12--4.14) of \citet{Lai:1993di}.  Hence, the mass-energy density and
pressure are determined as
\begin{align}
\epsilon(n,x)&  = n\big[m_n+E(n,x)/c^2\big]  \label{eq: energy-density} \,, \\
P(n,x) & = n^2{\frac{\partial E(n,x)}{\partial n}}
=\frac{2n}{3}T_n+\frac{n}{3}T_\e+n^2\left[V_{0}^{\prime}+V_{2}^{\prime} \,(1-2x)^2\right]  \label{eq: pressure} \,,
\end{align}
where
\begin{equation}
T_\e(n,x_\e)=\frac{3}{4}\hbar c(3\pi^2n)^{1/3}x_\e^{4/3}  \,, \label{eq: energy_e}
\end{equation}
is the energy per baryon of relativistic electrons.  Here, and in the following,
primes denote baryon number density $n$ derivatives (for example,
$V_{0}^{\prime}={\rm d} V_{0}/{{\rm d} n}$).  The  adiabatic sound speed
$c_\s^2$ is
\begin{multline}
c_\s^2 =\frac{\partial P}{\partial \epsilon}
=\frac{n}{\epsilon+P/c^2}\frac{\partial P}{\partial n}  \\
\quad =  \frac{n}{\epsilon+P/c^2}\left\{{\frac{10}{9}}T_n 
+{\frac{4}{9}}T_\e+2n\left[V_{0}^{\prime}+V_{2}^{\prime}\,(1-2x)^2\right]  \right\} \\
+\frac{n}{\epsilon+P/c^2}\left\{n^2\left[V_{0}^{\prime\prime}+V_{2}^{\prime\prime}\,(1-2x)^2\right] \right\} \,.
\label{eq: adiabatic_cs}
\end{multline}
The difference between $c_\s^2$ and $c_\e^2$ is given by
\begin{multline}
c_\s^2-c_\e^2 =\frac{n}{\epsilon+P/c^2}\left(\frac{\partial P}{\partial n}
-\frac{{\rm d} P}{{\rm d} n}\right)=-\frac{n}{\epsilon+P/c^2}
\left(\frac{\partial P}{\partial x}\right){\frac{{\rm d} x}{{\rm d} n}}\\
=-\frac{n^3}{\epsilon+P/c^2}\left[{\frac{\partial }{\partial n}}(\mu_\e+\mu_\p-\mu_\n)
\right]{\frac{{\rm d} x}{{\rm d} n}} \,.
\label{eq: delt_cs}
\end{multline}
From the beta equilibrium [i.e.\ Eq.~(\ref{eq: one})], we obtain
\begin{equation}
{\frac{{\rm d} x}{{\rm d} n}}= -\left[{\frac{\partial }{\partial n}}(\mu_\e+\mu_\p-\mu_\n) \right]\left[{\frac{\partial }{\partial x}}(\mu_\e+\mu_\p-\mu_\n) \right] ^{-1}\,.
 \label{eq: dx/dn}
\end{equation}
Finally, the difference between $c_\s^2$ and $c_\e^2$ can be represented as
\begin{equation}
c_\s^2-c_\e^2 = \frac{n^3}{\epsilon+P/c^2}\left[{\frac{\partial }{\partial n}}(\mu_\e+\mu_\p-\mu_\n) \right]^{2}\left[{\frac{\partial }{\partial x}}(\mu_\e+\mu_\p-\mu_\n) \right] ^{-1}\,.
\label{eq: Delt_cs}
\end{equation}

\begin{figure*}
    \centering 
    \includegraphics[width=12cm]{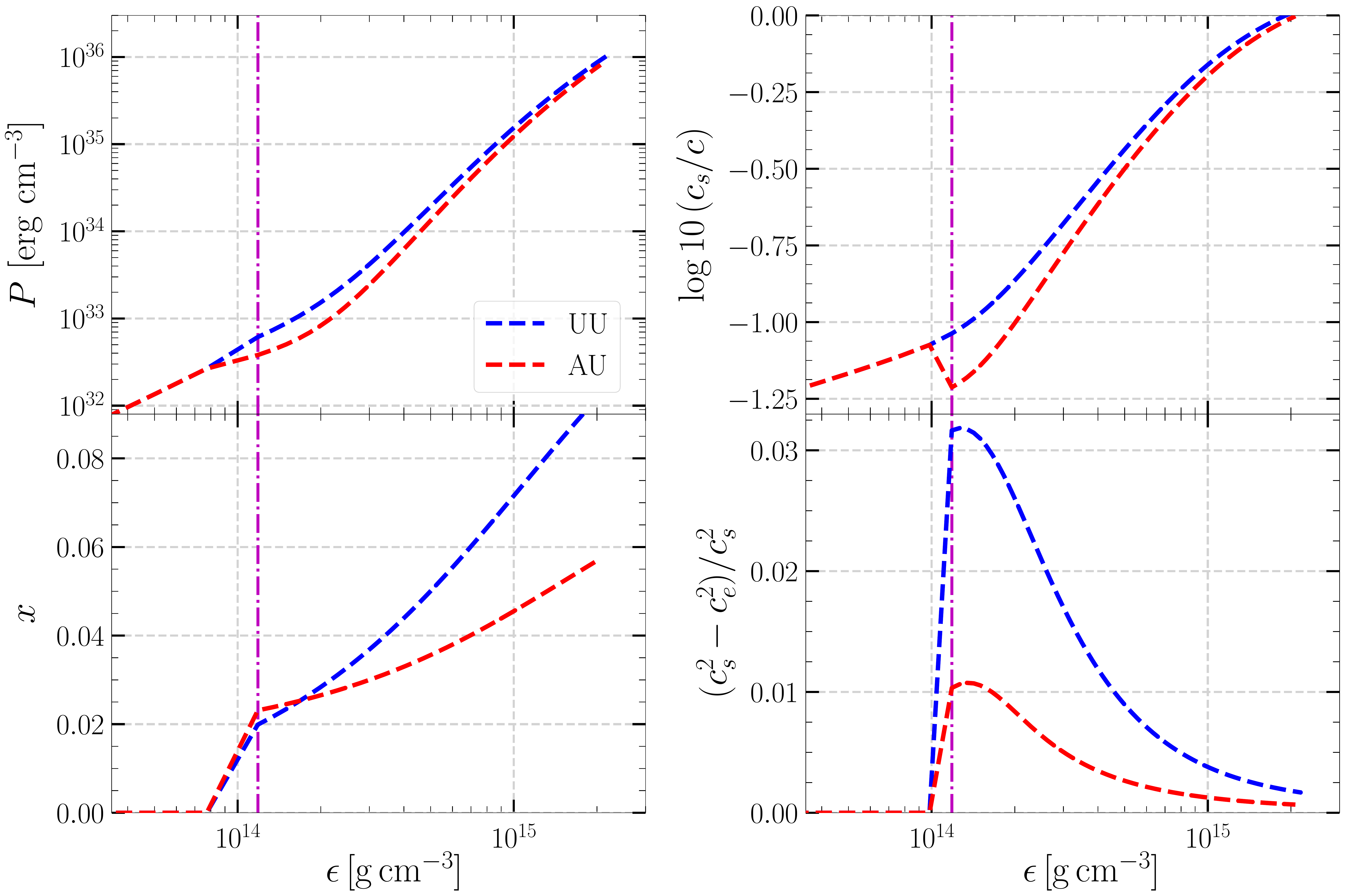}
    \caption{The left panels show the pressure $P$ (upper) and the proton fraction $x=n_\p/n$ (lower) versus the mass-energy density $\epsilon$ for representative EOS models AU and UU.  The right panels show the relation between the adiabatic sound speed $c_\s$ and the fractional difference between $c_\s^2$ and $c_\e^2$ versus the mass-energy density $\epsilon$. The purple dashed line is the mass-energy density $\epsilon = 0.07 \,\rm fm^{-3}$.}
    \label{fig: relations}
\end{figure*}

In the upper left  panel of Fig.~\ref{fig: relations}, we show the EOS models AU
and UU, which include below neutron-drip region \cite{BBP:1971} and the
lower-density crustal region \cite{BPS:1971}.  In the bottom left panel of
Fig.~\ref{fig: relations}, we show the relation between the proton fraction
$x=n_\p/n$ and the mass-energy density $\epsilon$.  One notices that the value
of $x$ of model UU is larger than that of model AU.  In the right panels of
Fig.~\ref{fig: relations}, we show the relation between  the adiabatic sound
speed $c_\s$ and the fractional difference between $c_\s^2$ and $c_\e^2$, as
functions of the mass-energy density.  Note that, in our work, we consider only
$g$-mode of the NS core, so we set $c_\s^2=c_\e^2$ in the lower-density region. 
As mentioned in Sec.~4 of \citet{Lai:1993di} that $c_\s^2=c_\e^2$ in the crustal
region indicates effectively suppressing the crustal $g$-mode while
concentrating on the core $g$-mode.

\begin{figure}
    \centering 
    \includegraphics[width=7cm]{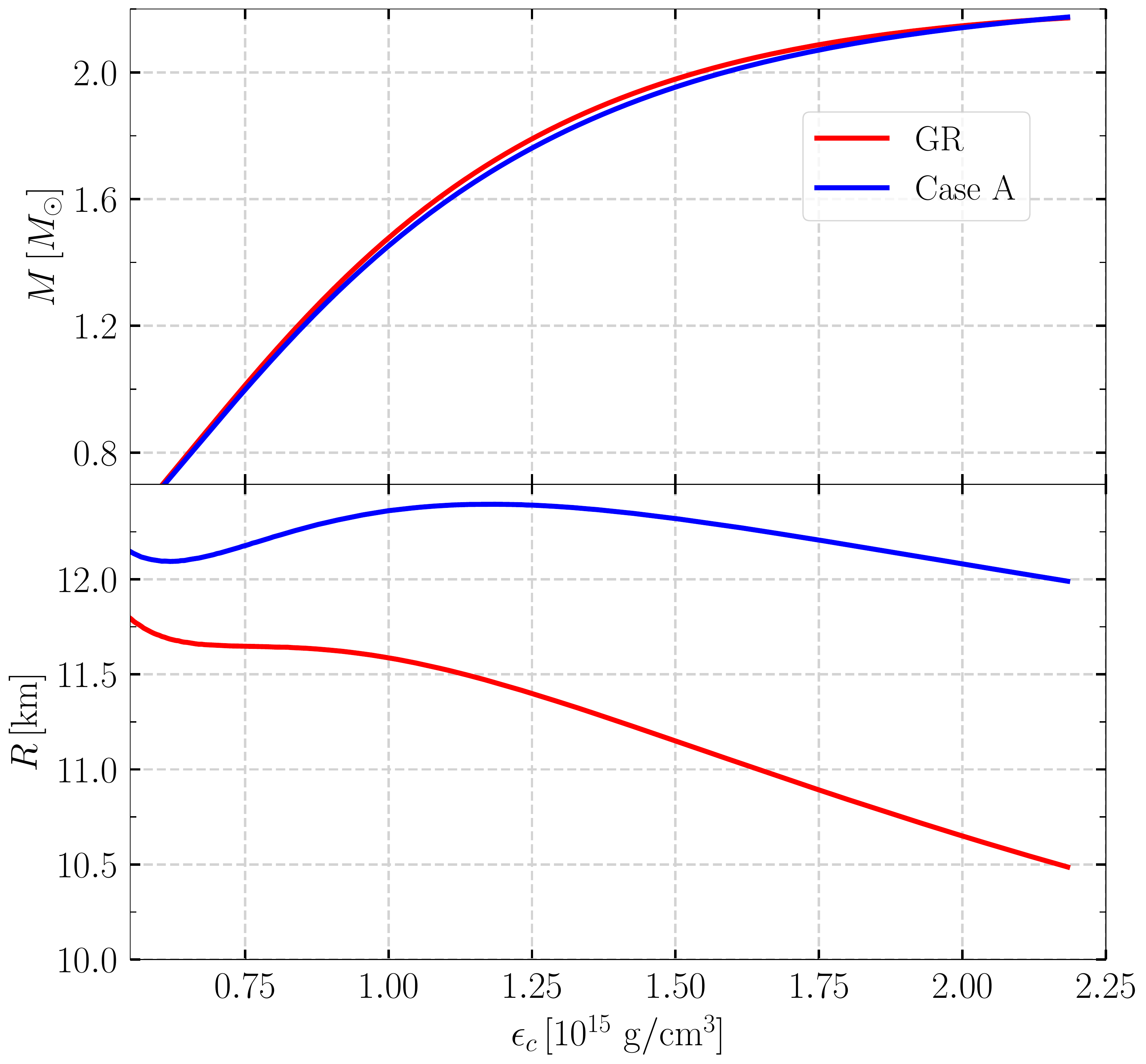}
    \caption{Mass and radius of model UU as a function of central density $\epsilon_{c}$. 
    The Case A and GR lines represent the background equations calculated by the Case A and TOV schemes, respectively.}
    \label{fig: MR_relation}
\end{figure}

As shown in Fig. \ref{fig: MR_relation}, the mass and radius of model UU are plotted against the
central density $\epsilon_{c}$. The Case A and GR lines represent the background equations calculated by the Case A and TOV schemes, respectively (see Table 1). We can see that the masses computed in the Case A formulation
can approximate well the GR solutions. The Case A formulation has absolute percentage differences
$5$--$17 \%$ for the radius of model UU. Note that the percentage difference of the stellar radii 
depends on the value of central density and the different EOS models. 
Detailed percentage differences of the stellar radii are illustrated in Appendix B of \citet{Tang:2021woo}.

\begin{figure}
    \centering 
    \includegraphics[width=7cm]{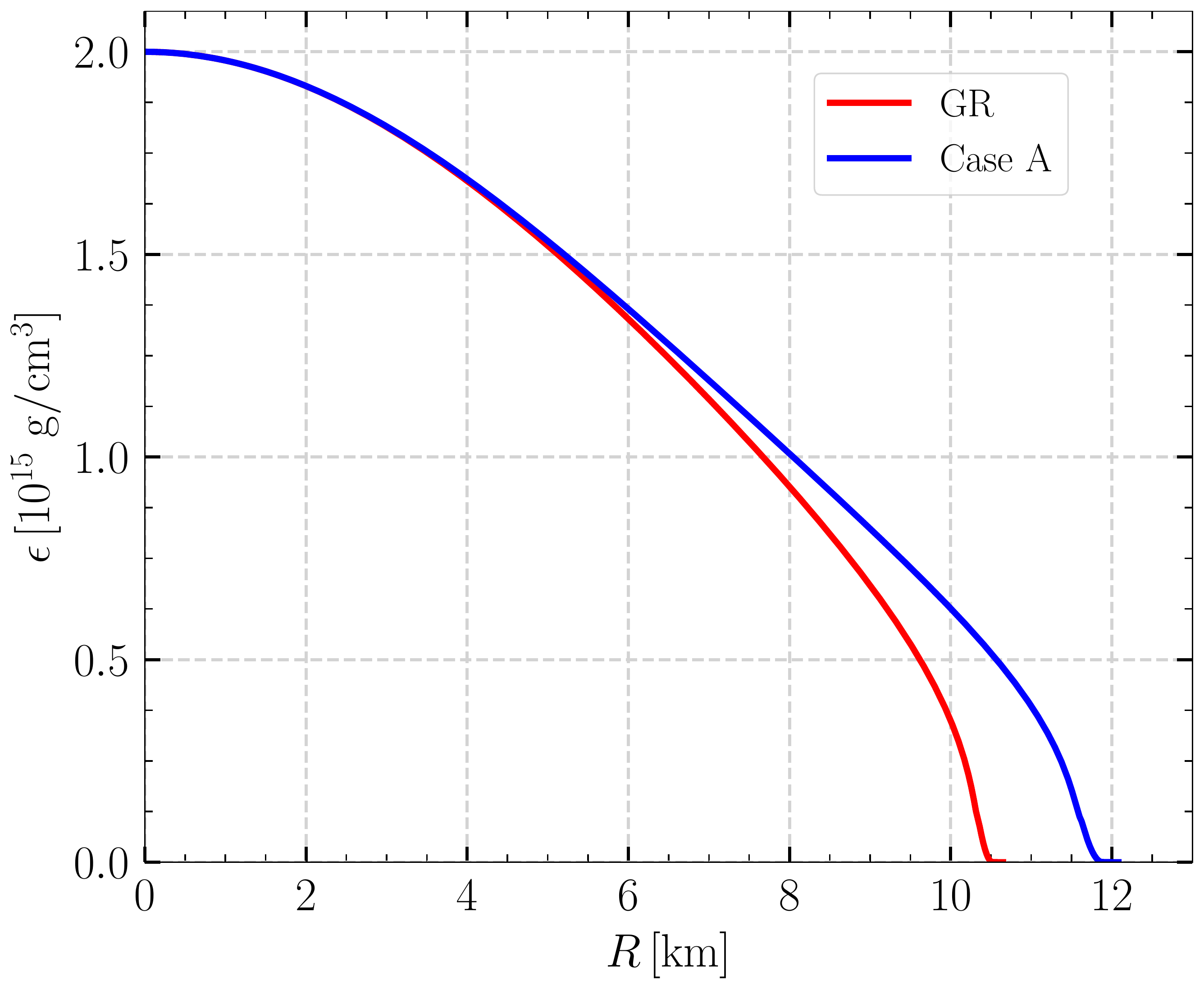}
    \caption{Comparison of the energy density profiles of the GR and Case A background solutions for model UU.}
    \label{fig: ER_relation}
\end{figure}

The energy density profiles of the GR and Case A schemes with 
central density $\epsilon=2.0 \times 10^{15} \rm g\,cm^{-3}$ for model UU is shown in Fig. \ref{fig: ER_relation}.
Compared with the GR solution, the Case A solution has a noticeable deviation only in the outer region of the surface. The total mass of the star is mainly determined by the high-density inner region.
The above results may explain  the fact that the total mass computed in the Case A formulation 
approximates well the GR solutions, though the radius has a large deviation.

Note that the rest-mass density $\rho$ appears in the background and 
perturbation equations in N and N+lapse schemes; the total energy density
$\epsilon$ and rest-mass density $\rho$ exhibit the background equations in Case
A and Case A+lapse schemes, but the rest-mass density $\rho$ appears in the
perturbation equations.
To compare with the results of \citet{Lai:1993di}, we
use the energy density $\epsilon$ to obtain the mass-radius relation, as well as
to solve perturbation equations. The difference between Case A and GR is
apparent, though much smaller than the difference between Newtonian gravity and
GR. Case A potential has captured some main effects from the full GR. As we
will see, the perturbation results will be even closer to that of GR than the
background results.

\citet{Lai:1993di} investigated $f$ and $g$ mode frequencies of EOS models AU
and UU with a given mass $M=1.4 \, M_{\odot}$\footnote{\citet{Lai:1993di} also
calculated models UT and UU2.  However, the maximum mass of the model UT does
not accord with the new observation results \cite{Antoniadis:2013pzd,
Fonseca:2021wxt}.  Also the model UU2 only considers the free $\n, \p,\e$
($V_{0}= V_{2}=0$).  We will not include the two EOSs in our calculations.}.
They found that the $f$-mode properties are very similar, due to the fact that
the two EOSs have similar bulk properties ($V_{0}$) for the nuclear matter. 
However, the properties of the $g$-mode are very different from models AU and
UU.  From the bottom right panel of Fig. \ref{fig: relations}, we find that the
value of $(c_\s^2-c_\e^2)/c_\s^2$ is different with increase of the energy
density.  These differences reflect the sensitive dependence of $g$-mode on the
nuclear matter's symmetry energy ($V_{2}$).

\begin{figure}
    \centering 
    \includegraphics[width=8cm]{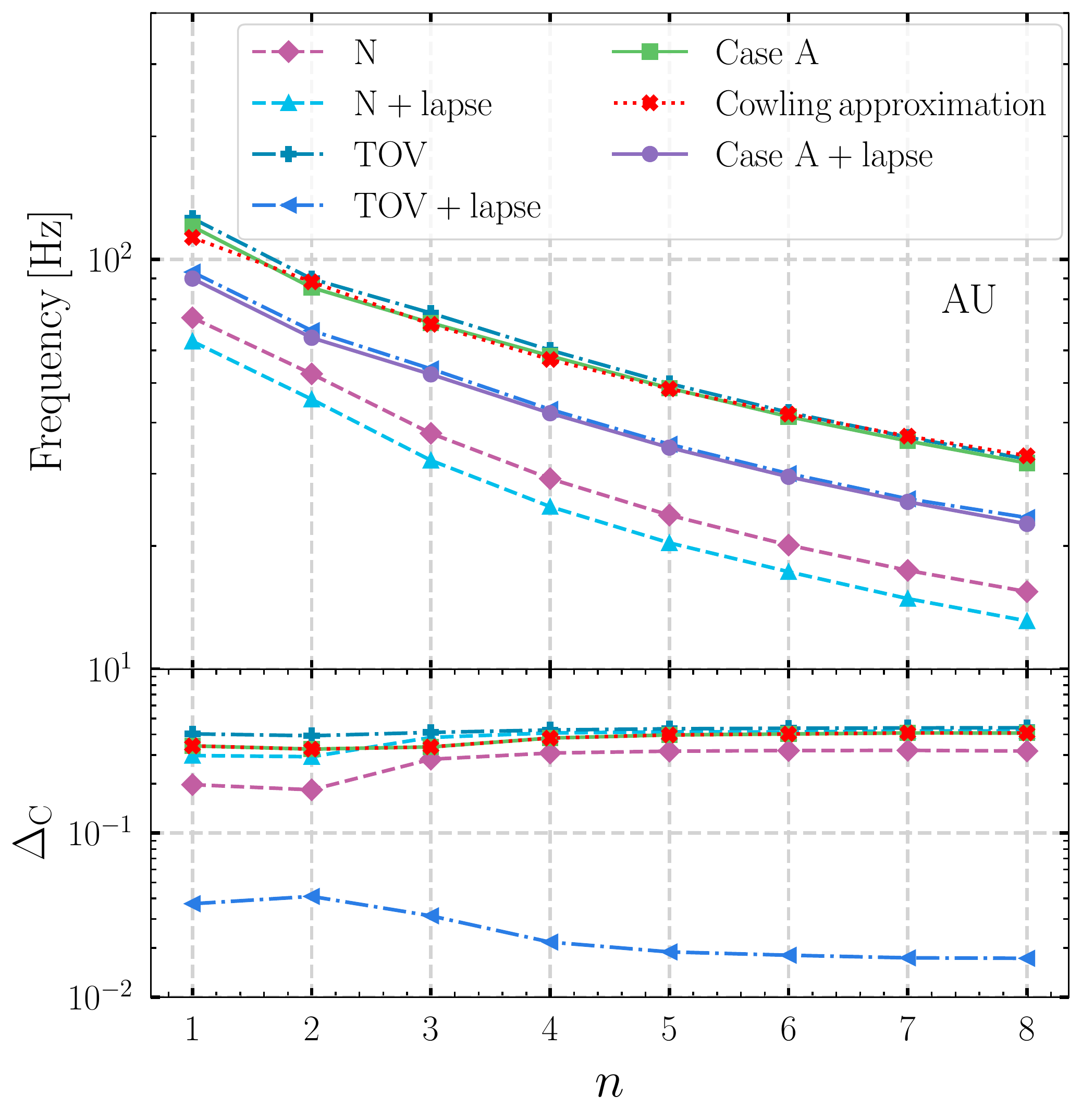}
    \caption{The $g$-mode frequencies for the EOS model AU, with a given mass
    $M=1.98 \, M_{\odot}$. The upper panel shows the frequencies of the first
    eight quadrupolar ($\ell=2$) $g$-modes with different schemes. 
    The lower panel shows the absolute fractional difference
    $\Delta_{\rm C}$ between our numerical results and the Case A+lapse scheme.
    } \label{fig: AU_frequency}
\end{figure}

\begin{figure}
    \centering 
    \includegraphics[width=8cm]{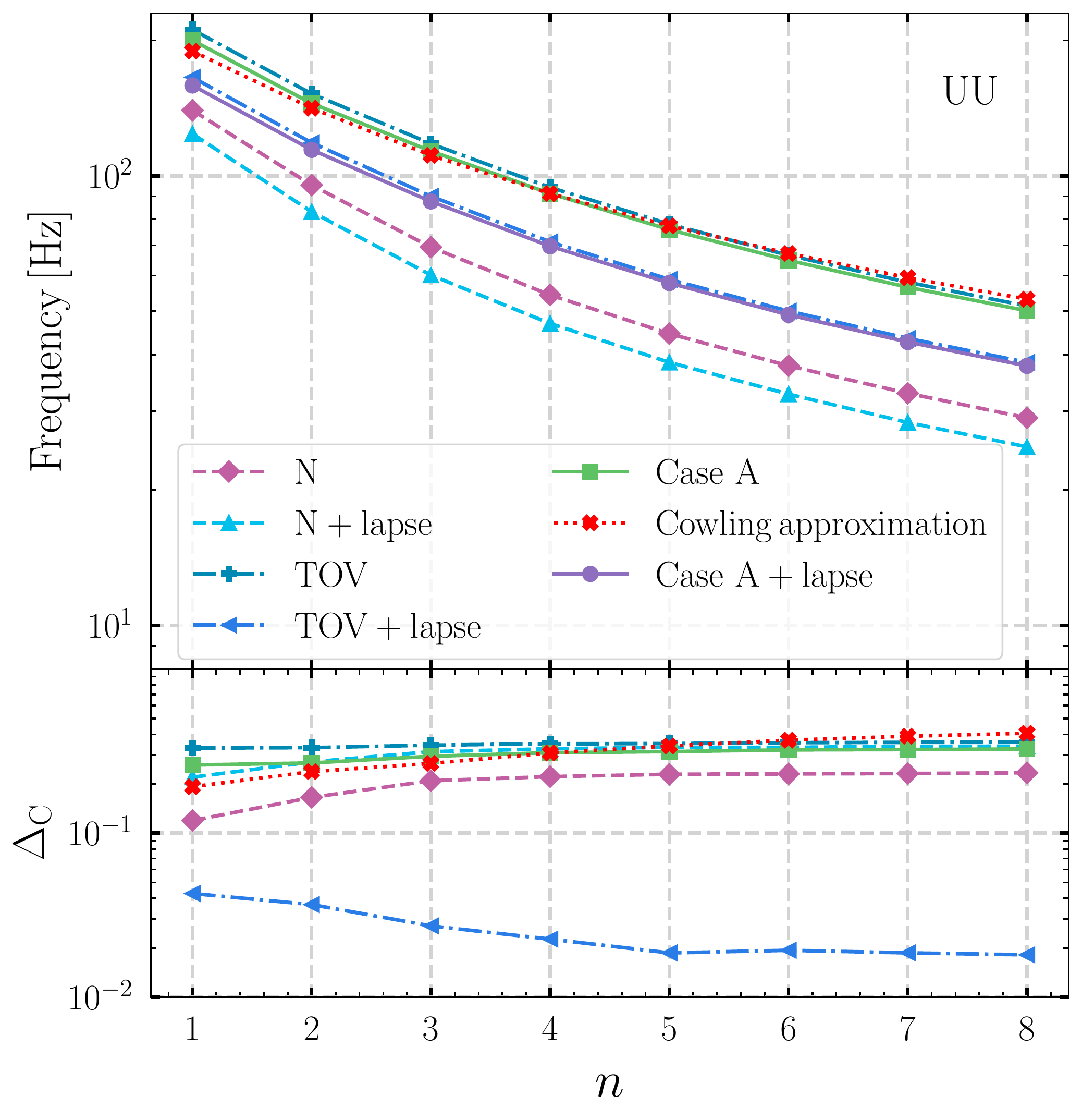}
    \caption{Same as Fig. \ref{fig: AU_frequency}, bur for the EOS model UU.}
    \label{fig: UU_frequency}
\end{figure}

In our study, we extend calculations in  \citet{Lai:1993di} by computing the
$g$-mode. We use the stars with a fixed mass $M=1.98 \, M_{\odot}$ as an
example.  In the upper panel of Fig. {\ref{fig: AU_frequency}}, we plot the
frequencies of the first eight quadrupolar $g$-mode for the EOS AU. The results
computed by all perturbation schemes are represented by different color lines in
Fig. {\ref{fig: AU_frequency}}. The lower panel of Fig. {\ref{fig:
AU_frequency}} shows the absolute fraction difference $\Delta_{\rm C}$ defined
by
\begin{equation}
\Delta_{\rm C} = \left\lvert\frac{f-f_\text{Case A+lapse}}{f_\text{Case A+lapse}}\right\rvert  \,,
\label{eq: error}
\end{equation}
where $f$ is the frequency of $g$-mode obtained by our perturbation schemes in
Table \ref{tab: schemes}.  According to the numerical results of non-radial
oscillation ($f$-mode) in \citet{Tang:2021woo}, the Case A+lapse scheme can 
approximate to about a few percents for a given mass $M=1.4 \, M_{\odot}$ in
full GR.  Hence, we use the results of Case A+lapse as the baseline in the case
of the composition gradient.  In particular, we found that the TOV+lapse scheme
can give a good approximation to the $g$-mode frequencies to a few percent
levels. Besides, the absolute percentage difference $\Delta_{\rm C}$ of the
TOV+lapse scheme decreases with increasing nodes.  We also plot the results of
frequencies of $g$-mode and the absolute percentage difference $\Delta_{\rm C}$
for the EOS UU in Fig. {\ref{fig: UU_frequency}}.  We seen similar properties of
$g$-mode, as the EOS AU in Fig. {\ref{fig: AU_frequency}}.

\subsection{Density discontinuity}\label{sec: density discontinuity}

In this subsection, we study the effect of  discontinuities at high density on
the oscillation spectrum of a NS.  We consider a simple polytropic EOS of the
form \cite{Finn:1987, McDermott:1990, Miniutti:2002bh}
\begin{equation}
 P = \left\{
      \begin{array}{rl}
            K\epsilon^{\gamma} \,,  \quad \quad \quad \quad & \epsilon > \epsilon_{\rm{d}} +\Delta\epsilon \,, \\
            \displaystyle{ K\left( 1 + \frac{\Delta \epsilon}{\epsilon_{\rm{d}}}
           \right)^{\gamma} \epsilon^{\gamma}} \,,             & \epsilon \leq \epsilon_{\rm{d}}     \,,  
           \end{array} \right.
           \label{eq: eos}  
\end{equation}
where the discontinuity of amplitude $\Delta\epsilon$ is located at a
mass-energy density $\epsilon_{\rm d}$. We study the properties of $g$-modes
with density discontinuity using the pseudo-Newtonian gravity schemes in
Table~\ref{tab: schemes}.

\begin{figure}
    \centering 
    \includegraphics[width=8cm]{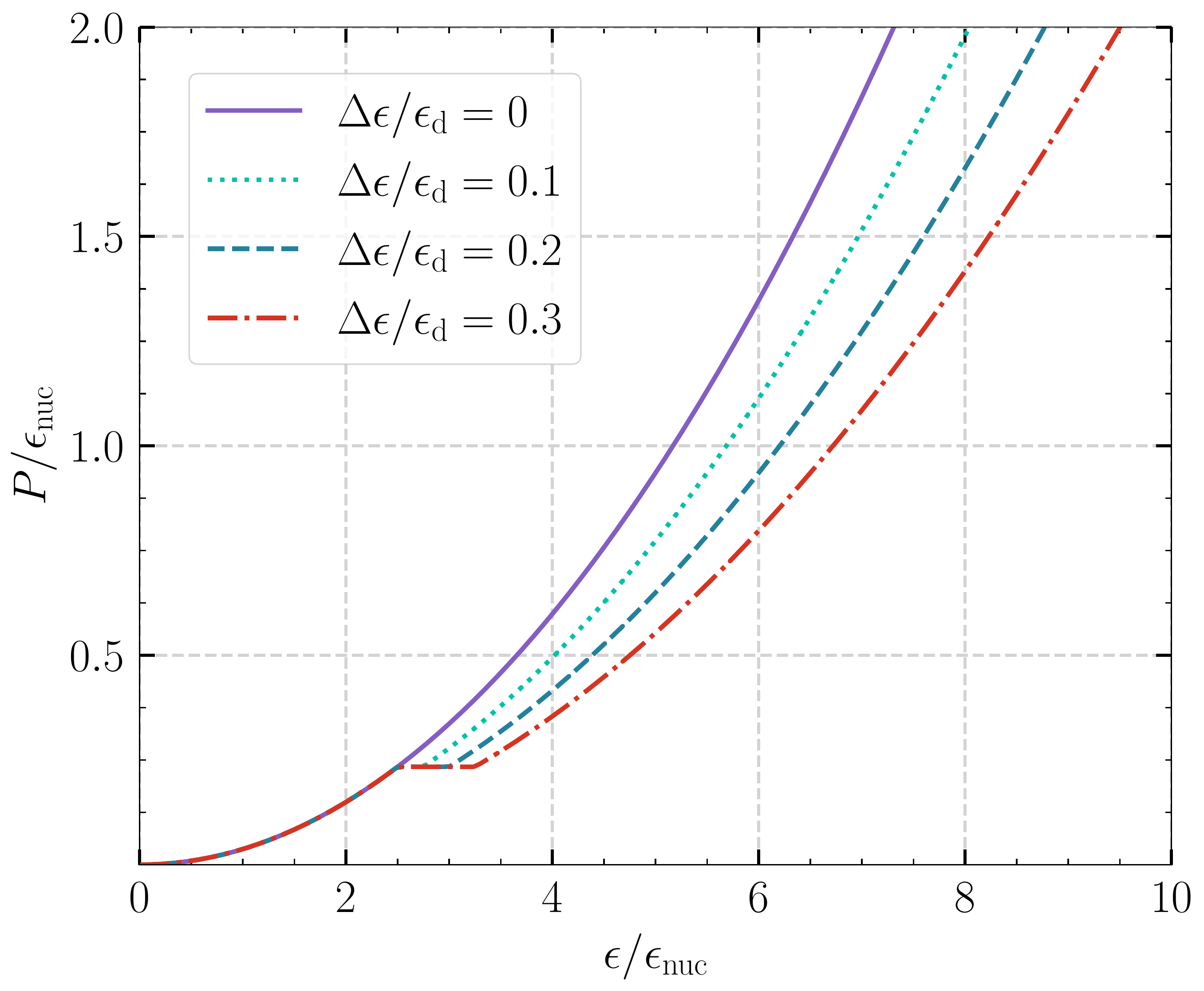}
    \caption{EOSs with density discontinuity,  for different values of $\Delta
    \epsilon/ \epsilon_{\rm {d}}$. The density and pressure are normalized by
    the standard nuclear density $\epsilon_{\rm nuc}=2.68\times 10^{14} \ \rm g
    \ cm^{-3}$.}
    \label{fig: density_EOS}
\end{figure}

\begin{figure*}
    \centering 
    \includegraphics[width=12cm]{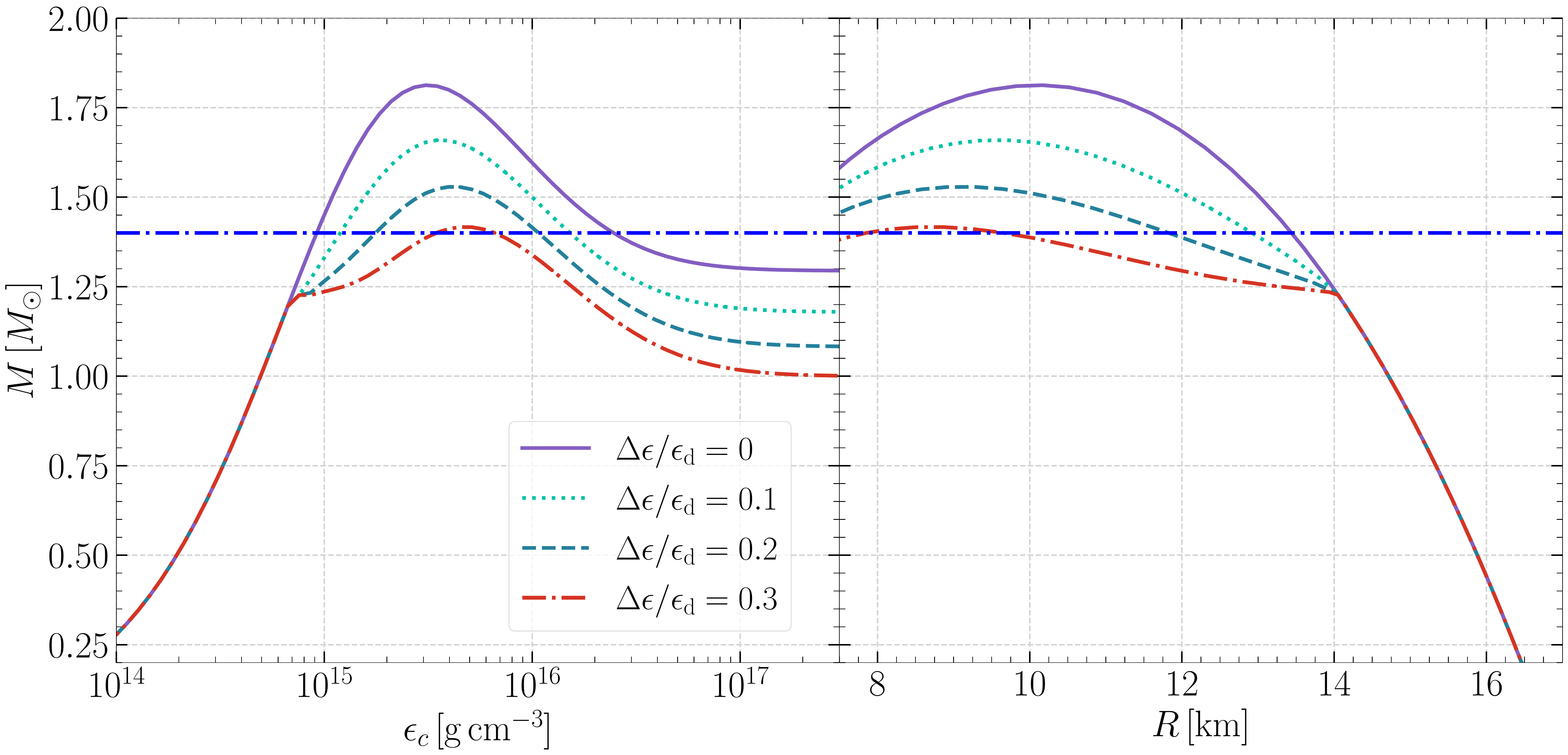}
    \caption{({\it Left}) The relation between the mass of NSs and central
    density $\epsilon_{c}$ for different values of $\Delta \epsilon/
    \epsilon_{\rm {d}}$. ({\it Right}) Mass and radius relation of NSs with the
    same value of  $\Delta \epsilon/ \epsilon_{\rm {d}}$. The horizontal blue line shows the mass
    $M=1.4 \, M_{\odot}$. } \label{fig: density_MR}
\end{figure*}

Now we have five parameters for a NS: the central density
$\epsilon_{c}$, the discontinuity of amplitude $\Delta\epsilon$, the critical
density $\epsilon_{\rm d}$, the polytropic index $\gamma$, and $K$.  To compared
with the results of non-radial oscillating relativistic stars  in the full
theory \cite[i.e.\ without the relativistic Cowling
approximation,][]{Miniutti:2002bh}, we  adopt the same parameters as
\citet{Miniutti:2002bh}: the polytropic index $\gamma=2$, $K=180\ \rm km^{2}$ 
for the NSs without discontinuity, and $K(1+\Delta \epsilon/\epsilon_{\rm
{d}})^{2}=180\ \rm km^{2}$ for the case with a discontinuity.  Some examples of
this EOS  are illustrated in Fig. {\ref{fig: density_EOS}}. 

\begin{figure*}
    \centering 
    \includegraphics[width=12cm]{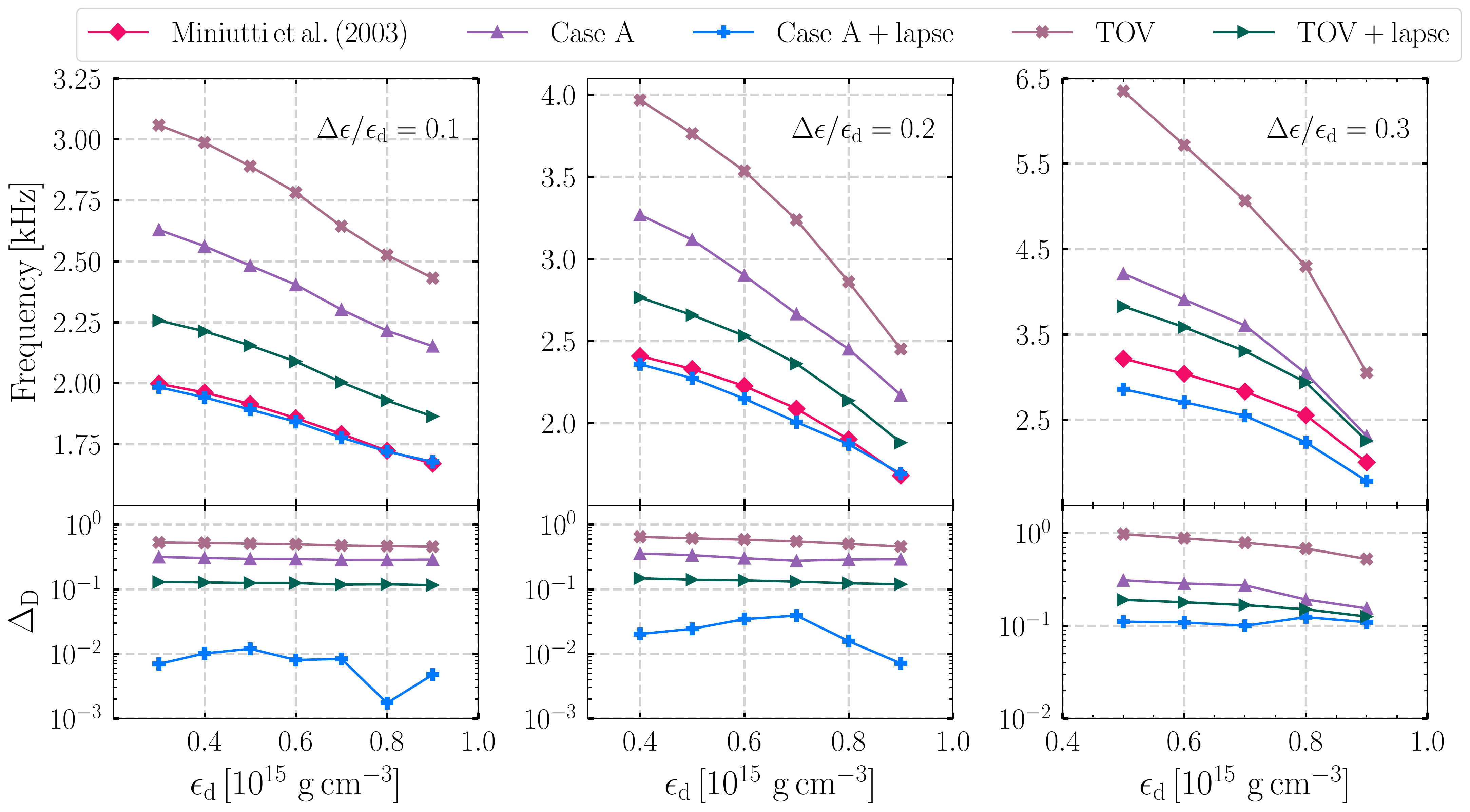}
    \caption{The top panels plot the frequency of the non-radial $f$-mode for
    different values of $\Delta \epsilon/ \epsilon_{\rm {d}}$ versus the density
    $\epsilon_{\rm d}$ for the four perturbation schemes. The bottom
    panels show the absolute fractional difference $\Delta_{\rm D}$ between our numerical
    results and the results of \citet{Miniutti:2002bh}.  We here consider stars with a fixed mass $M=1.4
    \, M_{\odot}$}. \label{fig: density_f_mode}
\end{figure*}

\begin{figure*}
    \centering 
    \includegraphics[width=12cm]{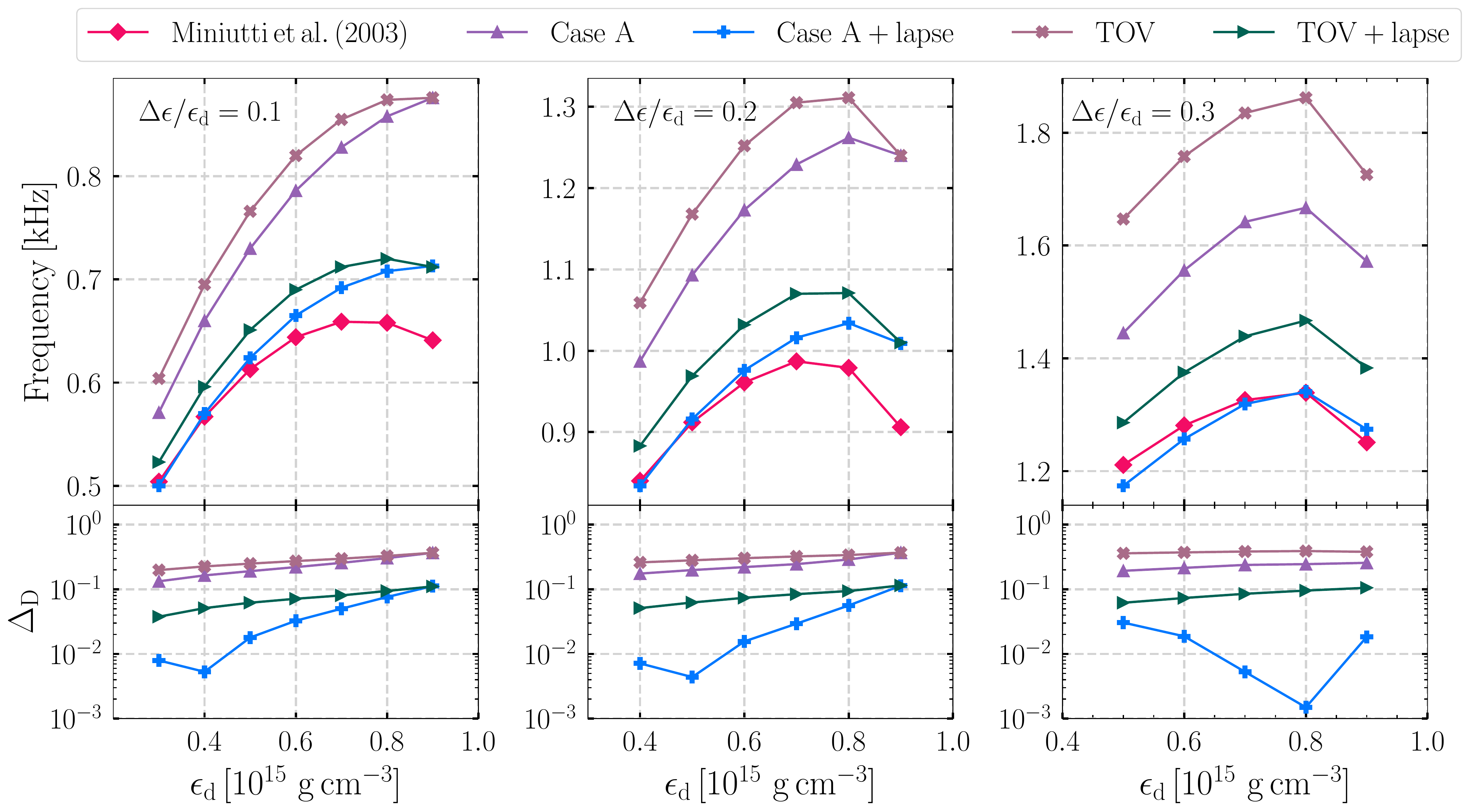}
    \caption{Same as Fig. {\ref{fig: density_f_mode}}, but for the $g$-mode
    frequencies.} \label{fig: density_g_mode}
\end{figure*}

In performing the calculation, boundary conditions must be specified at the
locations of the density discontinuities. \citet{Finn:1987} analyzed the jump
conditions of the perturbation variables with the Cowling approximation in
Newtonian gravity. Since the density is discontinuous, the perturbation
variables are discontinuous as well, and the differential equations (\ref{eq:
bc_p1}--\ref{eq: bc_p4}) require jump conditions in the discontinuity density,
denoted as [$\rho$]
\begin{align}
    & [U] = 0  \,, \\
    & [\delta\tilde{P} ]= g_{N} [\rho]U\,, \\
    & [\delta\tilde{\Phi}] = -4\pi[\rho]U  \,, \\
    & [\Psi ]=0 \,.
\end{align}
To compare with the results of \citet{Miniutti:2002bh}, we use the energy
density $\epsilon$ to solve perturbation equations.

In the left panel of Fig. {\ref{fig: density_MR}}, we show the mass $M$ versus
central density $\epsilon_{c}$ for each value of $\Delta \epsilon/ \epsilon_{\rm
{d}}$.  As $\Delta \epsilon/ \epsilon_{\rm {d}}$ gets larger, the maximum mass
decreases, and the stable region $\dd M/ \rm d\epsilon_{c}>0$ becomes narrower
and moves to a high-density region.  In this work, we study only stable NS
models with $\rm d M/ \rm d\epsilon_{c}>0$.  In our analysis, we fix the mass of
a NS to $M=1.4 \, M_{\odot}$ as an example.  In the right panel of Fig.
{\ref{fig: density_MR}},  we plot the mass-radius relation for NSs with and
without density discontinuity.  In both cases, we set the polytropic index
$\gamma=2$.  Comparing to  the same EOS for $\epsilon<\epsilon_{\rm {d}}$, we
adopt $K=180\ \rm km^{2}$ for the NS models without discontinuity, and
$K(1+\Delta \epsilon/ \epsilon_{\rm {d}})^{2}=180\ \rm km^{2}$ for the NS models
with discontinuity.  We find that the maximum mass is lower for the model with a
discontinuity.  Because the softening of EOS affected by the discontinuity.  NSs
with a discontinuity are more compact than those without discontinuity for a
fixed mass.

Now we will focus on the $\ell=2$ non-radial oscillation modes. In particular,
we consider the quadrupolar fundamental $f$-mode and gravity $g$-mode.  The
frequency versus density $\epsilon_{\rm {d}}$ for the fixed mass $M=1.4 \,
M_{\odot}$ is shown in the top panel of Fig. {\ref{fig: density_f_mode}}.  The
results computed by the four different perturbation schemes are represented by
different color lines in Fig. {\ref{fig: density_f_mode}}.  The GR curves in the
upper panel correspond to the results of full perturbation theory in GR
\cite{Miniutti:2002bh}.  Additionally, the absolute fraction difference
$\Delta_{\rm D}$ defined by 
\begin{equation}
\Delta_{\rm D} = \left\lvert\frac{f-f_\text{GR}}{f_\text{GR}}\right\rvert  \,,
\label{eq: error_D}
\end{equation}
is shown in the bottom panel of Fig. {\ref{fig: density_f_mode}}.  The frequency
of $f$-mode of the Case A+lapse scheme decreases with increasing density
$\epsilon_{\rm d}$, which is similar to the GR results in trend.  Again, the
Case A+lapse scheme is quite accurate for the frequency of the $f$-mode.  For
the $\Delta \epsilon/ \epsilon_{\rm {d}}=0.3$, the Case A+lapse scheme is not as
good as that of the $\Delta \epsilon/ \epsilon_{\rm {d}}=0.1, 0.2$ cases, but it
is still the best among the four perturbation schemes.  \citet{Tang:2021woo}
calculated $f$-mode using Newtonian, Newtonian+lapse, Case A, and  Case A+lapse
schemes.  They found that the Case A+lapse scheme performs much better and  can
reasonably approximate the $f$-mode frequency. 

\begin{figure*}
    \centering 
    \includegraphics[width=12cm]{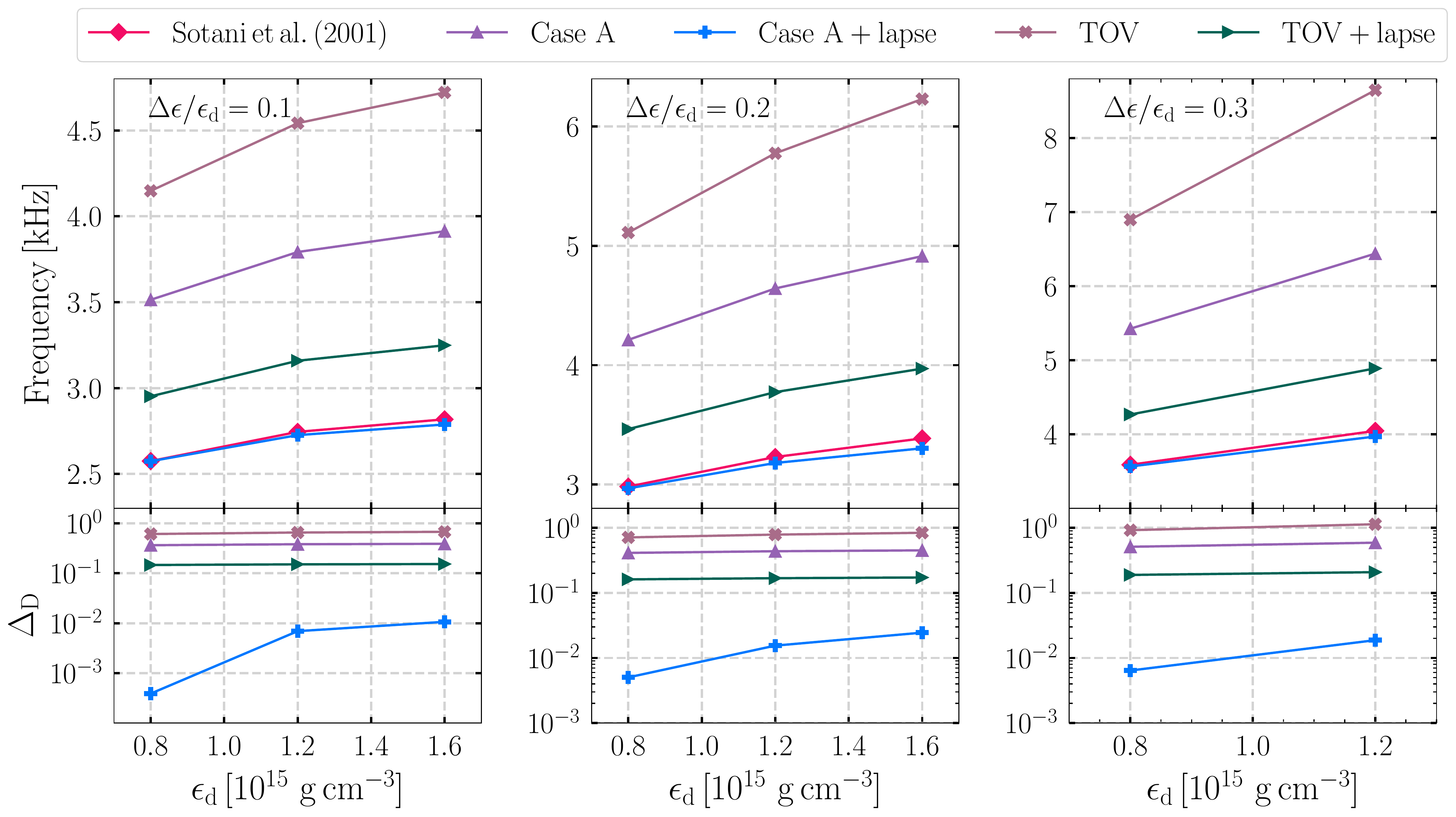}
    \caption{The top panels plot the frequency of the non-radial $f$-mode for
    different values of $\Delta \epsilon/ \epsilon_{\rm {d}}$ versus the density
    $\epsilon_{\rm d}$ for the four perturbation schemes. The bottom
    panels show the absolute fractional difference $\Delta_{\rm D}$ between our numerical
    results and  the results of \citet{Sotani:2001bb}.  
    We here consider stars with a fixed mass $M=1.2 \, M_{\odot}$}.
    \label{fig: density_f_mode_1.2}
\end{figure*}

\begin{figure*}
    \centering 
    \includegraphics[width=12cm]{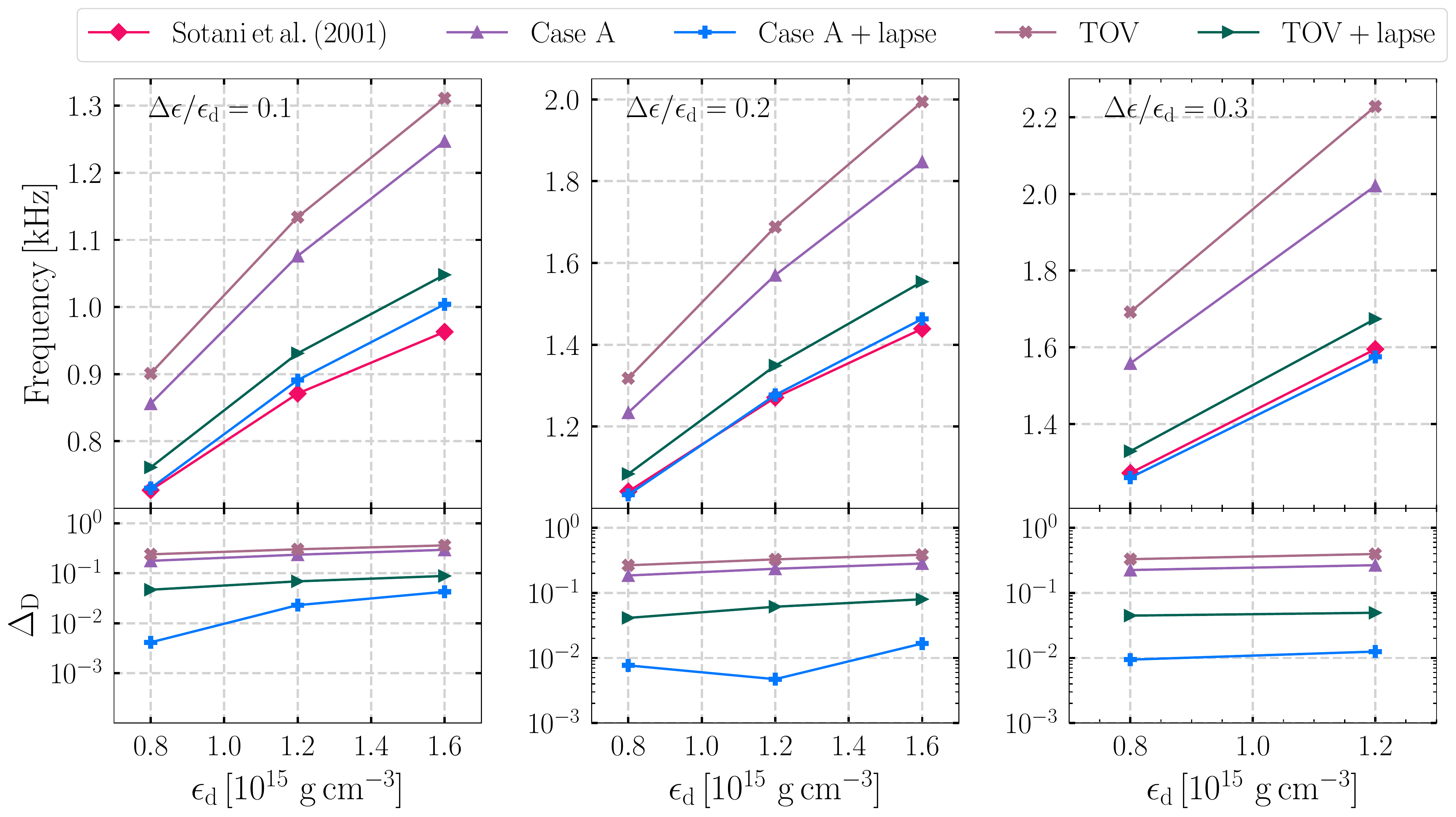}
    \caption{Same as Fig. {\ref{fig: density_f_mode_1.2}}, but for the $g$-mode
    frequencies.} \label{fig: density_g_mode_1.2}
\end{figure*}

 \begin{figure}
    \centering 
    \includegraphics[width=8cm]{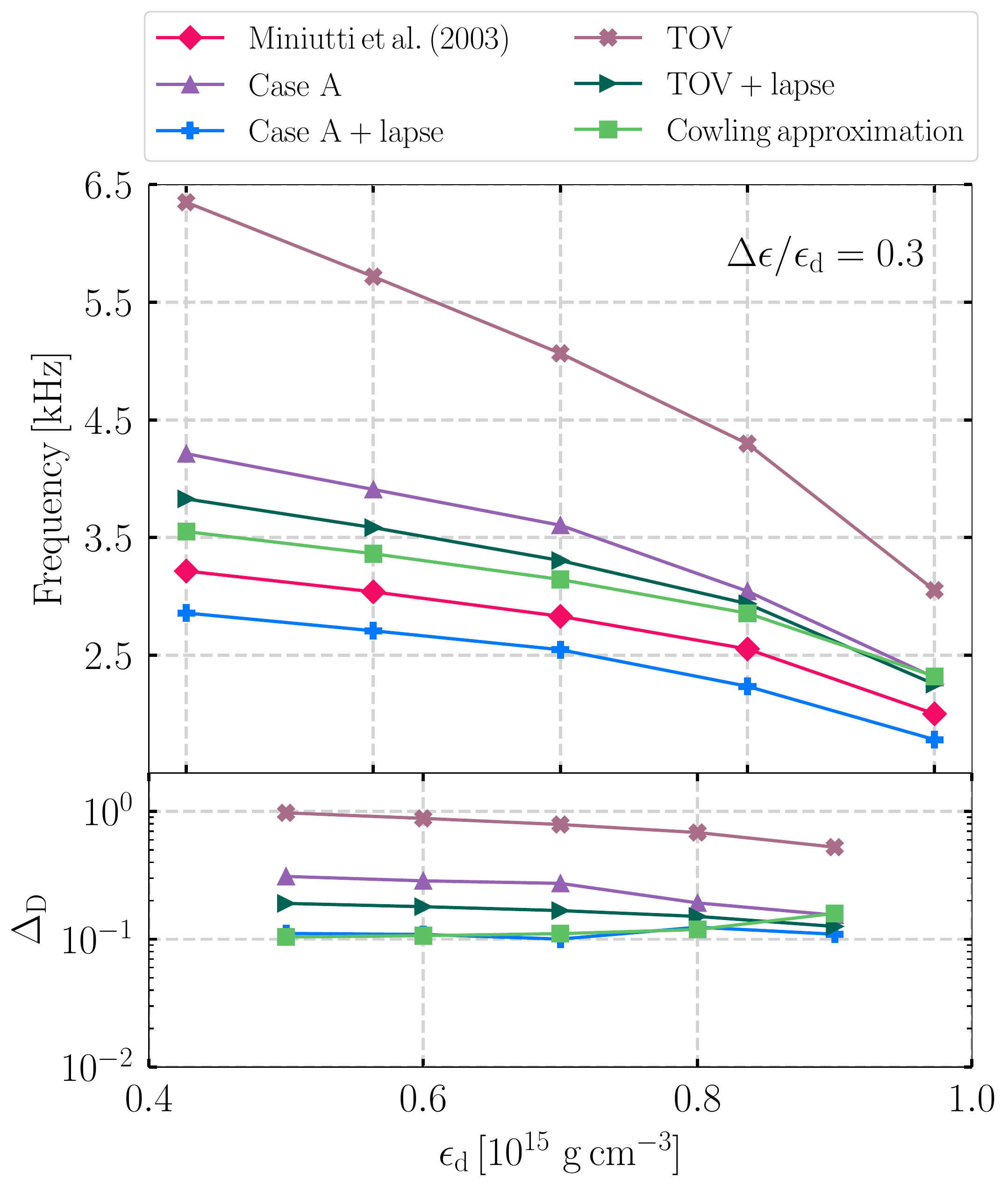}
    \caption{\Reply{ The top panel plots the frequency of the non-radial $f$-mode with $\Delta \epsilon/ \epsilon_{\rm {d}}=0.3$ versus the density
    $\epsilon_{\rm d}$ for the different perturbation schemes.  The bottom
    panel shows the absolute fractional difference $\Delta_{\rm D}$ between our numerical results and the results of \citet{Miniutti:2002bh}. Here we consider stars with a fixed mass $M=1.4
    \, M_{\odot}$.}} \label{fig: Cowling_f_mode}
\end{figure}

\begin{figure}
    \centering 
    \includegraphics[width=8cm]{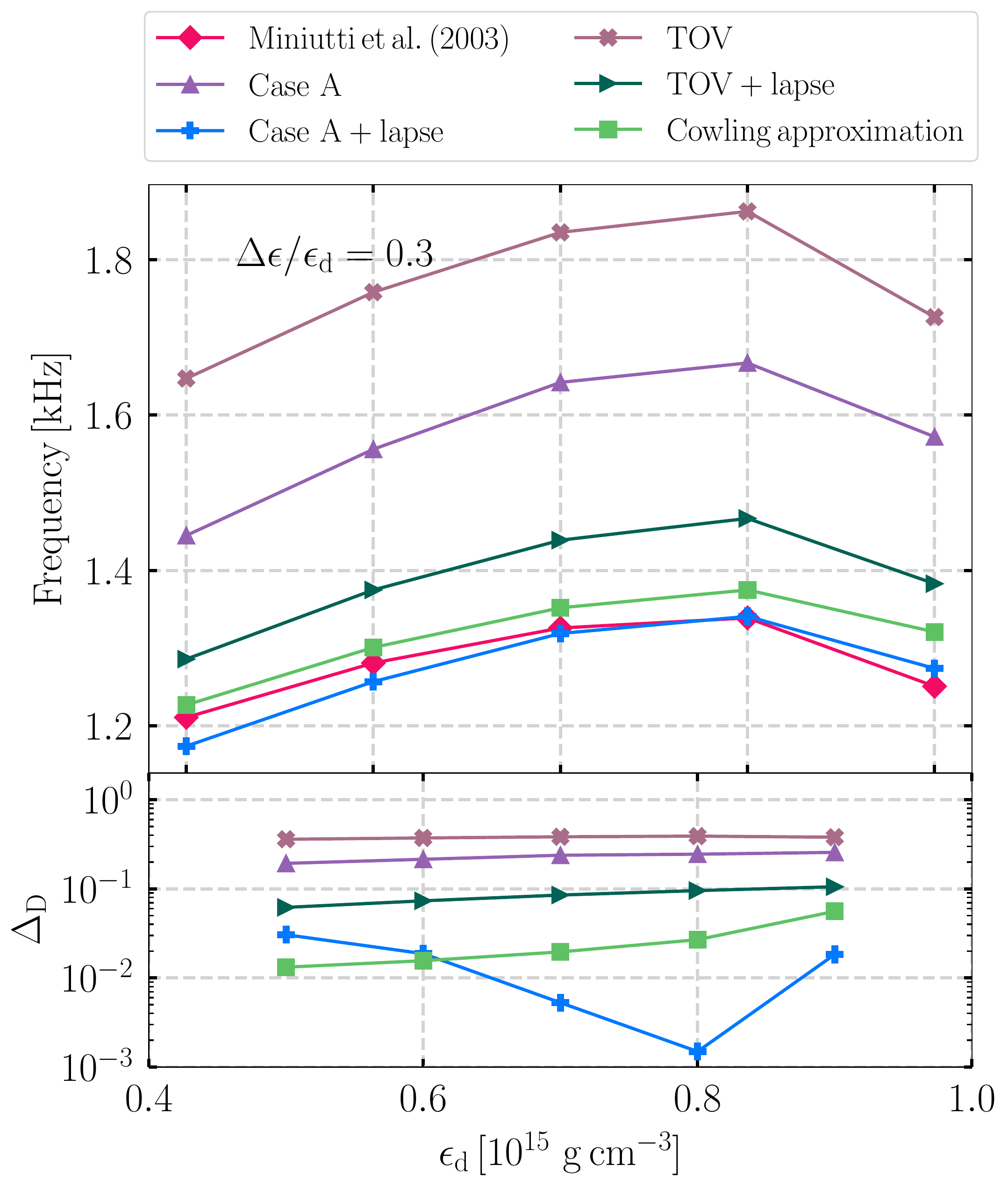}
    \caption{\Reply{Same as Fig. {\ref{fig: Cowling_f_mode}}, but for the $g$-mode
    frequencies.}} \label{fig: Cowling_g_mode}
\end{figure}

In the top panel of Fig. {\ref{fig: density_g_mode}}, we show the frequency of
$g$-mode as a function of the density $\epsilon_{\rm d}$ for the four schemes
and the results of \citet{Miniutti:2002bh}.
We also plot the results of $\Delta_{\rm D}$ for the four
schemes at the bottom of Fig. {\ref{fig: density_g_mode}}.  In particular, we
find that the Case A+lapse scheme can approximate the $g$-mode frequency of GR
reasonably well \cite{Miniutti:2002bh}. The percentage difference $\Delta_{\rm
D}$ of $g$-mode of the Case A+lapse scheme decreases with increasing $\Delta
\epsilon/ \epsilon_{\rm {d}}$.  The Case A+lapse scheme provides the best
approximation to the frequencies of $f$ and $g$ modes.  For the same central
density and discontinuity density, the radius of density discontinuity
$R_{\rm{d}}$ is larger than the radius $R$ of the Newtonian star.  Hence, we
ignore the N and N+lapse schemes of discontinuity $g$-mode in this work. 
Numerical results of the different schemes are given in Tables \ref{tab: f_mode}
and \ref{tab: g_mode}.

To prove that the pseudo-Newtonian treatments can
approximate well the GR solutions, we also calculate the $f$ and $g$ modes 
of mass $M=1.2 \, M_{\odot}$ for the different schemes.  
Detailed numerical results of the different schemes are given in Fig. {\ref{fig: density_f_mode_1.2}}, Fig. {\ref{fig: density_g_mode_1.2}}, as well as in 
Tables \ref{tab: f_mode_1.2} and \ref{tab: g_mode_1.2}.
We find that the pseudo-Newtonian gravity
can accurately describe the oscillation of the relativistic NSs constructed from an
EOS with a first-order phase transition.

For a given density $\epsilon_{\rm d}$ and $\Delta
\epsilon/ \epsilon_{\rm d}$, we show our numerical results for the frequencies
of $f$ and $g$ modes with four schemes and the GR scheme, where the GR results
were calculated by \citet{Miniutti:2002bh, Sotani:2001bb}. They integrated the equations describing the polar, 
non-radial perturbations of a non-rotating star as formulated by \citet{Lindblom:1983} and \citet{Detweiler:1985}.

\Reply{Finally, \citet{Sotani:2001bb} used the Cowling approximation to calculate the $f$ and $g$ modes and compared them to the results obtained from full GR. The results computed by the different perturbation schemes are represented by different colored lines in  Fig. {\ref{fig: Cowling_f_mode}} and Fig. {\ref{fig: Cowling_g_mode}}.
We can see that the Case A+lapse scheme provides the best
approximation to the frequency of $g$-mode when the central density increases.}

\begin{table*}
    \centering
    \caption{Comparison between the frequencies of $f$-mode (unit: Hz) of
    \citet{Miniutti:2002bh} and the different schemes in Table \ref{tab:
    schemes}, with a given mass $M=1.4 \, M_{\odot}$	and $\Gamma=2 $ with
    different central densities. The polytropic coefficient $K$ is $K(1+\Delta
    \epsilon/ \epsilon)^{2}=180\, \rm km^{2}$. }
        \renewcommand\arraystretch{1.3}
        \begin{tabular}{c c c c c c c }
        \hline
        $\epsilon_{\rm d}\ (\rm g\,cm^{-3})$ &  $\Delta \epsilon/ \epsilon_{\rm d}$ & \citet{Miniutti:2002bh} & Case A & Case A+lapse &TOV & TOV+lapse \\
        \hline
        --                            & 0.0 & 1666 & 2144 & 1673 & 2423 & 1863\\
        \hline
        $3\times10^{14}$ & 0.1 & 1998 & 2629 & 1984 & 3058 & 2257\\
        $4\times10^{14}$ & 0.1 & 1962 & 2562 & 1942 & 2987 & 2213\\
        $5\times10^{14}$ & 0.1 & 1915 & 2482 & 1892 & 2890 & 2155\\
        $6\times10^{14}$ & 0.1 & 1857 & 2404 & 1842 & 2782 & 2089\\
        $7\times10^{14}$ & 0.1 & 1792 & 2302 & 1777 & 2644 & 2004\\
        $8\times10^{14}$ & 0.1 & 1723 & 2215 & 1720 & 2526 & 1929\\
        $9\times10^{14}$ & 0.1 & 1670 & 2152 & 1678 & 2431 & 1864\\
        \hline
        $4\times10^{14}$ & 0.2 & 2408 & 3269 & 2359 & 3968 & 2765\\
        $5\times10^{14}$ & 0.2 & 2330 & 3117 & 2273 & 3764 & 2658\\
        $6\times10^{14}$ & 0.2 & 2226 & 2901 & 2149 & 3536 & 2532\\
        $7\times10^{14}$ & 0.2 & 2088 & 2665 & 2006 & 3238 & 2362\\
        $8\times10^{14}$ & 0.2 & 1901 & 2451 & 1871 & 2860 & 2137\\
        $9\times10^{14}$ & 0.2 & 1680 & 2171 & 1692 & 2451 & 1881\\
        \hline
        $5\times10^{14}$ & 0.3 & 3216 & 4213 & 2859 & 6350 & 3829\\
        $6\times10^{14}$ & 0.3 & 3039 & 3909 & 2708 & 5718 & 3585\\
        $7\times10^{14}$ & 0.3 & 2831 & 3605 & 2547 & 5066 & 3305\\
        $8\times10^{14}$ & 0.3 & 2553 & 3044 & 2236 & 4298 & 2938\\
        $9\times10^{14}$ & 0.3 & 2002 & 2311 & 1783 & 3053 & 2254\\
        \hline
    \end{tabular}
    \label{tab: f_mode}
\end{table*}

\begin{table*}
    \centering
    \caption{Same as Table \ref{tab: f_mode}, but for the $g$-mode frequencies.}
        \renewcommand\arraystretch{1.3}
        \begin{tabular}{c c c c c c c }
        \hline
        $\epsilon_{\rm d}\ (\rm g\,cm^{-3})$ &  $\Delta \epsilon/ \epsilon_{\rm d}$ & \citet{Miniutti:2002bh} & Case A & Case A+lapse & TOV & TOV+lapse \\
        \hline
        --                            & 0.0 & --    & --       & --      & --      & -- \\
        \hline
        $3\times10^{14}$ & 0.1 & 504 & 571 & 500 & 604 & 523\\
        $4\times10^{14}$ & 0.1 & 567 & 660 & 570 & 695 & 596\\
        $5\times10^{14}$ & 0.1 & 613 & 730 & 624 & 766 & 651\\
        $6\times10^{14}$ & 0.1 & 644 & 786 & 665 & 820 & 690\\
        $7\times10^{14}$ & 0.1 & 659 & 828 & 692 & 855 & 712\\
        $8\times10^{14}$ & 0.1 & 658 & 858 & 708 & 874 & 720\\
        $9\times10^{14}$ & 0.1 & 641 & 876 & 713 & 876 & 712\\
        \hline
        $4\times10^{14}$ & 0.2 & 840 & 987 & 834 & 1059 & 883\\
        $5\times10^{14}$ & 0.2 & 912 & 1093 & 916 & 1168 & 969\\
        $6\times10^{14}$ & 0.2 & 961 & 1173 & 976 & 1252 & 1032\\
        $7\times10^{14}$ & 0.2 & 987 & 1229 & 1016 & 1305 & 1070\\
        $8\times10^{14}$ & 0.2 & 979 & 1262 & 1034 & 1311 & 1071\\
        $9\times10^{14}$ & 0.2 & 906 & 1240 & 1009 & 1240 & 1010\\
        \hline
        $5\times10^{14}$ & 0.3 & 1211 & 1445 & 1174 & 1647 & 1286\\
        $6\times10^{14}$ & 0.3 & 1281 & 1556 & 1257 & 1758 & 1375\\
        $7\times10^{14}$ & 0.3 & 1326 & 1642 & 1319 & 1835 & 1439\\
        $8\times10^{14}$ & 0.3 & 1339 & 1667 & 1341 & 1862 & 1467\\
        $9\times10^{14}$ & 0.3 & 1251 & 1572 & 1274 & 1726 & 1383\\
        \hline
    \end{tabular}
    \label{tab: g_mode}
\end{table*}


\begin{table*}
    \centering
    \caption{Comparison between the frequencies of $f$-mode (unit: Hz) of
    \citet{Sotani:2001bb} and the different schemes, with a given mass $M=1.2 \, M_{\odot}$	
    for different central densities.}
        \renewcommand\arraystretch{1.3}
        \begin{tabular}{c c c c c c c }
        \hline\hline
        $\epsilon_{\rm d}\ (\rm g\,cm^{-3})$ &  $\Delta \epsilon/ \epsilon_{\rm d}$ & \citet{Sotani:2001bb} & Case A & Case A+lapse &TOV & TOV+lapse \\
        \hline
        $8\times10^{14}$    & 0.1 & 2575 & 3514 & 2574 & 4148& 2952\\
        $1.2\times10^{15}$ & 0.1 & 2745 & 3792 & 2726 & 4542 & 3159\\
        $1.6\times10^{15}$ & 0.1 & 2818 & 3913 & 2788 & 4721 & 3249\\
        \hline
        $8\times10^{14}$    & 0.2 & 2982 & 4212 & 2967 & 5112 & 3464\\
        $1.2\times10^{15}$ & 0.2 & 3230 & 4643 & 3180 & 5776 & 3774\\
        $1.6\times10^{15}$ & 0.2 & 3386 & 4914 & 3303 & 6231 & 3971\\
        \hline
        $8\times10^{14}$    & 0.3 & 3588 & 5426 & 3565 & 6898 & 4267\\
        $1.2\times10^{15}$ & 0.3 & 4046 & 6440 & 3970 & 6647 & 4890\\
        \hline
    \end{tabular}
    \label{tab: f_mode_1.2}
\end{table*}

\begin{table*}
    \centering
    \caption{Same as Table \ref{tab: f_mode_1.2}, but for the $g$-mode frequencies.}
        \renewcommand\arraystretch{1.3}
        \begin{tabular}{c c c c c c c }
        \hline\hline
        $\epsilon_{\rm d}\ (\rm g\,cm^{-3})$ &  $\Delta \epsilon/ \epsilon_{\rm d}$ & \citet{Sotani:2001bb} & Case A & Case A+lapse &TOV & TOV+lapse \\
        \hline
        $8\times10^{14}$    & 0.1 & 727  & 856   & 730  & 901    & 761\\
        $1.2\times10^{15}$ & 0.1 & 871  & 1076 & 891  & 1134  & 931\\
        $1.6\times10^{15}$ & 0.1 & 963  & 1247 &1004 & 1311  & 1048\\
        \hline
        $8\times10^{14}$    & 0.2 & 1041 & 1234 & 1033 & 1318 & 1084\\
        $1.2\times10^{15}$ & 0.2 & 1271 & 1570 & 1277 & 1688 & 1349\\
        $1.6\times10^{15}$ & 0.2 & 1439 & 1847 & 1463 & 1994 & 1554\\
        \hline
        $8\times10^{14}$    & 0.3 & 1272 & 1558 & 1260 & 1692 & 1329\\
        $1.2\times10^{15}$ & 0.3 & 1595 & 2021 & 1575 & 2228 & 1674\\
        \hline
    \end{tabular}
    \label{tab: g_mode_1.2}
\end{table*}

\section{Conclusions}\label{sec: conculsion}

In light of new observations, oscillating modes of NSs are of particular
interests to the physics and astrophysics communities in recent years.  In this
work, we have investigated the properties of the gravity $g$-mode for NSs in the
framework of pseudo-Newtonian gravity.  \citet{Tang:2021woo} have investigated
barotropic oscillations ($\Gamma_{1}=\gamma$ and  the Schwarzschild discriminant
$A = 0$).  We extended the work and have studied the $g$-mode of NSs with the
same polytropic EOS model.  We find that, the $g$-mode frequencies increase with
increasing adiabatic index, which indicates that the buoyancy becomes much
larger.

A deeper understanding of the oscillation of NSs, which could be associated with
emitted gravitational waves, requires an analysis of both the state and
composition of the NS matter.  We considered the case of the composition
gradient,  and have extended calculations in \citet{Lai:1993di} to compute the
$g$-mode.  The value of $(c_\s^2-c_\e^2)/c_\s^2$ is different when the energy
density increases.  In particular, these differences reflect the sensitive 
dependence of $g$-mode on the nuclear matter's symmetry energy [$V_{2}$ in
Eq.~(\ref{eq:Ennx}))].  Note that the tidal deformability of binary NSs  appears
to be related to the dominant oscillation frequency of the post-merger remnant
\cite{Bernuzzi:2015rla}.  The impact of thermal and rotational effects can
provide simple arguments that help explain the result
\cite{Chakravarti:2019sdc}.  More recently,  \citet{Andersson:2022cax} consider
the dynamic tides of NSs to build the structure NSs in the framework of
post-Newtonian gravity.  We may expect using the pseudo-Newtonian gravity to
study the resonant oscillations and tidal response in coalescing binary NSs in
the future.

We considered a phase transition occurring in the inner core of NSs, which could
be associated with a density discontinuity.  Phase transition would produce a
softening of EOSs, leading to more compact NSs.  Using the different schemes, we
have calculated the frequencies of $f$ and $g$ modes for the $\ell=2$ component.
Compared to the results of GR \cite{Miniutti:2002bh, Sotani:2001bb}, the Case A+lapse scheme
can approximate the $f$-mode frequency very well. The absolute percentage
difference $\Delta_{\rm D}$ ranges from $0.01$ to $0.1$ percent.  In particular,
we find that the Case A+lapse scheme also can approximate the $g$-mode frequency
of GR  reasonably well \cite{Miniutti:2002bh, Sotani:2001bb}.  The percentage difference
$\Delta_{\rm D}$ of $g$-mode of the Case A+lapse scheme decreases with
increasing $\Delta \epsilon/ \epsilon_{\rm {d}}$ in our model. 

The existence of a possible hadron-quark phase transition in the central regions
of NSs is associated with the appearance of $g$-mode, which is extremely
important as they could signal the presence of a pure quark matter core in the
center of NSs~\cite{Orsaria:2019ftf}.  Our findings suggest that the
pseudo-Newtonian gravity, with much less computational efforts than the full GR,
can accurately study the oscillation of the relativistic NSs constructed from an
EOS with a first-order phase transition.  Observations of $g$-mode frequencies
with density discontinuity may thus be interpreted as a possible hint of the
first-order phase transition in the core of NSs. Lastly, our work also provides
more confidence in using the pseudo-Newtonian gravity in the simulations of
CCSNs, thus reducing the computational cost significantly.

\citet{McDermott:1988} investigated the non-radial oscillation of NSs using Cowling approximation 
and discussed the different damping mechanisms. \citet{Reisenegger:1992} considered the $g$-mode induced by composition (proton-to-neutron ratio) gradient in the cores of NSs and discussed damping mechanisms. They also estimated damping rates for the core $g$-modes.
\citet{Cutler:1990} assessed the accuracy of Cowling eigenfunctions and found that the relativistic Cowling approximation by \citet{McDermott:1983} accurately predicts frequencies and eigenfunctions.
\citet{Chugunov:2011dc} calculated the non-radial oscillations of superfluid non-rotating stars.
An approximate decoupling of
equations describing the oscillation modes of superfluid and normal fluid has been studied by \citet{Gusakov:2011}.
Further, \citet{Gusakov:2012zx} developed an approximate method to determine the eigenfrequencies and eigenfunctions of an oscillating superfluid NS.
In this work, we
found that the $g$-mode frequencies in one of the pseudo-Newtonian treatments can
approximate remarkably well the GR solutions than the relativistic Cowling approximation.
Hence, we may conjecture that  
the eigenfunction and dissipation of the pseudo-Newtonian treatments are also more accurate than the relativistic Cowling approximation. 
Based on the approximate method of \citet{Gusakov:2012zx}, 
one can calculate the eigenfunctions and the different damping mechanisms in future studies.

\begin{acknowledgments}
We thank the anonymous referee for helpful comments, and Zexin Hu and Yacheng Kang for the helpful discussions.  This work was
supported by the National SKA Program of China (2020SKA0120300, 2020SKA0120100),
the National Natural Science Foundation of China (11975027, 11991053), the
National Key R\&D Program of China (2017YFA0402602), the Max Planck Partner
Group Program funded by the Max Planck Society, and the High-Performance
Computing Platform of Peking University. 
\end{acknowledgments}

\bibliography{ref}

\begin{thebibliography}{62}%
\makeatletter
\providecommand \@ifxundefined [1]{%
 \@ifx{#1\undefined}
}%
\providecommand \@ifnum [1]{%
 \ifnum #1\expandafter \@firstoftwo
 \else \expandafter \@secondoftwo
 \fi
}%
\providecommand \@ifx [1]{%
 \ifx #1\expandafter \@firstoftwo
 \else \expandafter \@secondoftwo
 \fi
}%
\providecommand \natexlab [1]{#1}%
\providecommand \enquote  [1]{``#1''}%
\providecommand \bibnamefont  [1]{#1}%
\providecommand \bibfnamefont [1]{#1}%
\providecommand \citenamefont [1]{#1}%
\providecommand \href@noop [0]{\@secondoftwo}%
\providecommand \href [0]{\begingroup \@sanitize@url \@href}%
\providecommand \@href[1]{\@@startlink{#1}\@@href}%
\providecommand \@@href[1]{\endgroup#1\@@endlink}%
\providecommand \@sanitize@url [0]{\catcode `\\12\catcode `\$12\catcode
  `\&12\catcode `\#12\catcode `\^12\catcode `\_12\catcode `\%12\relax}%
\providecommand \@@startlink[1]{}%
\providecommand \@@endlink[0]{}%
\providecommand \url  [0]{\begingroup\@sanitize@url \@url }%
\providecommand \@url [1]{\endgroup\@href {#1}{\urlprefix }}%
\providecommand \urlprefix  [0]{URL }%
\providecommand \Eprint [0]{\href }%
\providecommand \doibase [0]{https://doi.org/}%
\providecommand \selectlanguage [0]{\@gobble}%
\providecommand \bibinfo  [0]{\@secondoftwo}%
\providecommand \bibfield  [0]{\@secondoftwo}%
\providecommand \translation [1]{[#1]}%
\providecommand \BibitemOpen [0]{}%
\providecommand \bibitemStop [0]{}%
\providecommand \bibitemNoStop [0]{.\EOS\space}%
\providecommand \EOS [0]{\spacefactor3000\relax}%
\providecommand \BibitemShut  [1]{\csname bibitem#1\endcsname}%
\let\auto@bib@innerbib\@empty
\bibitem [{\citenamefont {{Andersson}}(2019)}]{Andersson:2019}%
  \BibitemOpen
  \bibfield  {author} {\bibinfo {author} {\bibfnamefont {N.}~\bibnamefont
  {{Andersson}}},\ }\href@noop {} {\emph {\bibinfo {title} {{Gravitational-Wave
  Astronomy: Exploring the Dark Side of the Universe}}}}\ (\bibinfo
  {publisher} {Oxford University Press},\ \bibinfo {year} {2019})\BibitemShut
  {NoStop}%
\bibitem [{\citenamefont {{Cowling}}(1941)}]{Cowling:1941}%
  \BibitemOpen
  \bibfield  {author} {\bibinfo {author} {\bibfnamefont {T.~G.}\ \bibnamefont
  {{Cowling}}},\ }\bibfield  {title} {\bibinfo {title} {{The non-radial
  oscillations of polytropic stars}},\ }\href
  {https://doi.org/10.1093/mnras/101.8.367} {\bibfield  {journal} {\bibinfo
  {journal} {Mon. Not. Roy. Astron. Soc.}\ }\textbf {\bibinfo {volume} {101}},\
  \bibinfo {pages} {367} (\bibinfo {year} {1941})}\BibitemShut {NoStop}%
\bibitem [{\citenamefont {Marek}\ \emph {et~al.}(2006)\citenamefont {Marek},
  \citenamefont {Dimmelmeier}, \citenamefont {Janka}, \citenamefont {Muller},\
  and\ \citenamefont {Buras}}]{Marek:2005if}%
  \BibitemOpen
  \bibfield  {author} {\bibinfo {author} {\bibfnamefont {A.}~\bibnamefont
  {Marek}}, \bibinfo {author} {\bibfnamefont {H.}~\bibnamefont {Dimmelmeier}},
  \bibinfo {author} {\bibfnamefont {H.~T.}\ \bibnamefont {Janka}}, \bibinfo
  {author} {\bibfnamefont {E.}~\bibnamefont {Muller}},\ and\ \bibinfo {author}
  {\bibfnamefont {R.}~\bibnamefont {Buras}},\ }\bibfield  {title} {\bibinfo
  {title} {{Exploring the relativistic regime with Newtonian hydrodynamics: An
  Improved effective gravitational potential for supernova simulations}},\
  }\href {https://doi.org/10.1051/0004-6361:20052840} {\bibfield  {journal}
  {\bibinfo  {journal} {A\&A}\ }\textbf {\bibinfo {volume} {445}},\ \bibinfo
  {pages} {273} (\bibinfo {year} {2006})},\ \Eprint
  {https://arxiv.org/abs/astro-ph/0502161} {arXiv:astro-ph/0502161}
  \BibitemShut {NoStop}%
\bibitem [{\citenamefont {Mueller}\ \emph {et~al.}(2008)\citenamefont
  {Mueller}, \citenamefont {Dimmelmeier},\ and\ \citenamefont
  {Mueller}}]{Mueller:2008it}%
  \BibitemOpen
  \bibfield  {author} {\bibinfo {author} {\bibfnamefont {B.}~\bibnamefont
  {Mueller}}, \bibinfo {author} {\bibfnamefont {H.}~\bibnamefont
  {Dimmelmeier}},\ and\ \bibinfo {author} {\bibfnamefont {E.}~\bibnamefont
  {Mueller}},\ }\bibfield  {title} {\bibinfo {title} {{Exploring the
  relativistic regime with Newtonian hydrodynamics: II. An effective
  gravitational potential for rapid rotation}},\ }\href
  {https://doi.org/10.1051/0004-6361:200809609} {\bibfield  {journal} {\bibinfo
   {journal} {A\&A}\ }\textbf {\bibinfo {volume} {489}},\ \bibinfo {pages}
  {301} (\bibinfo {year} {2008})},\ \Eprint {https://arxiv.org/abs/0802.2459}
  {arXiv:0802.2459 [astro-ph]} \BibitemShut {NoStop}%
\bibitem [{\citenamefont {Yakunin}\ \emph {et~al.}(2015)\citenamefont {Yakunin}
  \emph {et~al.}}]{Yakunin:2015wra}%
  \BibitemOpen
  \bibfield  {author} {\bibinfo {author} {\bibfnamefont {K.~N.}\ \bibnamefont
  {Yakunin}} \emph {et~al.},\ }\bibfield  {title} {\bibinfo {title}
  {{Gravitational wave signatures of ab initio two-dimensional core collapse
  supernova explosion models for 12\textendash{}25 M$_\odot$ stars}},\ }\href
  {https://doi.org/10.1103/PhysRevD.92.084040} {\bibfield  {journal} {\bibinfo
  {journal} {Phys. Rev. D}\ }\textbf {\bibinfo {volume} {92}},\ \bibinfo
  {pages} {084040} (\bibinfo {year} {2015})},\ \Eprint
  {https://arxiv.org/abs/1505.05824} {arXiv:1505.05824 [astro-ph.HE]}
  \BibitemShut {NoStop}%
\bibitem [{\citenamefont {Morozova}\ \emph {et~al.}(2018)\citenamefont
  {Morozova}, \citenamefont {Radice}, \citenamefont {Burrows},\ and\
  \citenamefont {Vartanyan}}]{Morozova:2018glm}%
  \BibitemOpen
  \bibfield  {author} {\bibinfo {author} {\bibfnamefont {V.}~\bibnamefont
  {Morozova}}, \bibinfo {author} {\bibfnamefont {D.}~\bibnamefont {Radice}},
  \bibinfo {author} {\bibfnamefont {A.}~\bibnamefont {Burrows}},\ and\ \bibinfo
  {author} {\bibfnamefont {D.}~\bibnamefont {Vartanyan}},\ }\bibfield  {title}
  {\bibinfo {title} {{The gravitational wave signal from core-collapse
  supernovae}},\ }\href {https://doi.org/10.3847/1538-4357/aac5f1} {\bibfield
  {journal} {\bibinfo  {journal} {Astrophys. J.}\ }\textbf {\bibinfo {volume}
  {861}},\ \bibinfo {pages} {10} (\bibinfo {year} {2018})},\ \Eprint
  {https://arxiv.org/abs/1801.01914} {arXiv:1801.01914 [astro-ph.HE]}
  \BibitemShut {NoStop}%
\bibitem [{\citenamefont {O'Connor}\ \emph {et~al.}(2018)\citenamefont
  {O'Connor} \emph {et~al.}}]{OConnor:2018sti}%
  \BibitemOpen
  \bibfield  {author} {\bibinfo {author} {\bibfnamefont {E.}~\bibnamefont
  {O'Connor}} \emph {et~al.},\ }\bibfield  {title} {\bibinfo {title} {{Global
  Comparison of Core-Collapse Supernova Simulations in Spherical Symmetry}},\
  }\href {https://doi.org/10.1088/1361-6471/aadeae} {\bibfield  {journal}
  {\bibinfo  {journal} {J. Phys. G}\ }\textbf {\bibinfo {volume} {45}},\
  \bibinfo {pages} {104001} (\bibinfo {year} {2018})},\ \Eprint
  {https://arxiv.org/abs/1806.04175} {arXiv:1806.04175 [astro-ph.HE]}
  \BibitemShut {NoStop}%
\bibitem [{\citenamefont {O'Connor}\ and\ \citenamefont
  {Couch}(2018)}]{OConnor:2015rwy}%
  \BibitemOpen
  \bibfield  {author} {\bibinfo {author} {\bibfnamefont {E.~P.}\ \bibnamefont
  {O'Connor}}\ and\ \bibinfo {author} {\bibfnamefont {S.~M.}\ \bibnamefont
  {Couch}},\ }\bibfield  {title} {\bibinfo {title} {{Two Dimensional
  Core-Collapse Supernova Explosions Aided by General Relativity with
  Multidimensional Neutrino Transport}},\ }\href
  {https://doi.org/10.3847/1538-4357/aaa893} {\bibfield  {journal} {\bibinfo
  {journal} {Astrophys. J.}\ }\textbf {\bibinfo {volume} {854}},\ \bibinfo
  {pages} {63} (\bibinfo {year} {2018})},\ \Eprint
  {https://arxiv.org/abs/1511.07443} {arXiv:1511.07443 [astro-ph.HE]}
  \BibitemShut {NoStop}%
\bibitem [{\citenamefont {Zha}\ \emph {et~al.}(2020)\citenamefont {Zha},
  \citenamefont {O'Connor}, \citenamefont {Chu}, \citenamefont {Lin},\ and\
  \citenamefont {Couch}}]{Zha:2020gjw}%
  \BibitemOpen
  \bibfield  {author} {\bibinfo {author} {\bibfnamefont {S.}~\bibnamefont
  {Zha}}, \bibinfo {author} {\bibfnamefont {E.~P.}\ \bibnamefont {O'Connor}},
  \bibinfo {author} {\bibfnamefont {M.-C.}\ \bibnamefont {Chu}}, \bibinfo
  {author} {\bibfnamefont {L.-M.}\ \bibnamefont {Lin}},\ and\ \bibinfo {author}
  {\bibfnamefont {S.~M.}\ \bibnamefont {Couch}},\ }\bibfield  {title} {\bibinfo
  {title} {{Gravitational-Wave Signature of a First-Order Quantum
  Chromodynamics Phase Transition in Core-Collapse Supernovae}},\ }\href
  {https://doi.org/10.1103/PhysRevLett.125.051102} {\bibfield  {journal}
  {\bibinfo  {journal} {Phys. Rev. Lett.}\ }\textbf {\bibinfo {volume} {125}},\
  \bibinfo {pages} {051102} (\bibinfo {year} {2020})},\ \bibinfo {note}
  {[Erratum: Phys.Rev.Lett. 127, 219901(E) (2021)]},\ \Eprint
  {https://arxiv.org/abs/2007.04716} {arXiv:2007.04716 [astro-ph.HE]}
  \BibitemShut {NoStop}%
\bibitem [{\citenamefont {Tang}\ and\ \citenamefont
  {Lin}(2022)}]{Tang:2021woo}%
  \BibitemOpen
  \bibfield  {author} {\bibinfo {author} {\bibfnamefont {Y.-T.}\ \bibnamefont
  {Tang}}\ and\ \bibinfo {author} {\bibfnamefont {L.-M.}\ \bibnamefont {Lin}},\
  }\bibfield  {title} {\bibinfo {title} {{Neutron star oscillations in
  pseudo-Newtonian gravity}},\ }\href {https://doi.org/10.1093/mnras/stab3687}
  {\bibfield  {journal} {\bibinfo  {journal} {Mon. Not. Roy. Astron. Soc.}\
  }\textbf {\bibinfo {volume} {510}},\ \bibinfo {pages} {3629} (\bibinfo {year}
  {2022})},\ \Eprint {https://arxiv.org/abs/2112.09474} {arXiv:2112.09474
  [astro-ph.HE]} \BibitemShut {NoStop}%
\bibitem [{\citenamefont {{Reisenegger}}\ and\ \citenamefont
  {{Goldreich}}(1992)}]{Reisenegger:1992}%
  \BibitemOpen
  \bibfield  {author} {\bibinfo {author} {\bibfnamefont {A.}~\bibnamefont
  {{Reisenegger}}}\ and\ \bibinfo {author} {\bibfnamefont {P.}~\bibnamefont
  {{Goldreich}}},\ }\bibfield  {title} {\bibinfo {title} {{A New Class of
  g-Modes in Neutron Stars}},\ }\href {https://doi.org/10.1086/171645}
  {\bibfield  {journal} {\bibinfo  {journal} {Astrophys. J.}\ }\textbf
  {\bibinfo {volume} {395}},\ \bibinfo {pages} {240} (\bibinfo {year}
  {1992})}\BibitemShut {NoStop}%
\bibitem [{\citenamefont {{McDermott}}\ \emph {et~al.}(1983)\citenamefont
  {{McDermott}}, \citenamefont {{van Horn}},\ and\ \citenamefont
  {{Scholl}}}]{McDermott:1983}%
  \BibitemOpen
  \bibfield  {author} {\bibinfo {author} {\bibfnamefont {P.~N.}\ \bibnamefont
  {{McDermott}}}, \bibinfo {author} {\bibfnamefont {H.~M.}\ \bibnamefont {{van
  Horn}}},\ and\ \bibinfo {author} {\bibfnamefont {J.~F.}\ \bibnamefont
  {{Scholl}}},\ }\bibfield  {title} {\bibinfo {title} {{Nonradial g-mode
  oscillations of warm neutron stars}},\ }\href
  {https://doi.org/10.1086/161006} {\bibfield  {journal} {\bibinfo  {journal}
  {Astrophys. J.}\ }\textbf {\bibinfo {volume} {268}},\ \bibinfo {pages} {837}
  (\bibinfo {year} {1983})}\BibitemShut {NoStop}%
\bibitem [{\citenamefont {{McDermott}}\ \emph {et~al.}(1988)\citenamefont
  {{McDermott}}, \citenamefont {{van Horn}},\ and\ \citenamefont
  {{Hansen}}}]{McDermott:1988}%
  \BibitemOpen
  \bibfield  {author} {\bibinfo {author} {\bibfnamefont {P.~N.}\ \bibnamefont
  {{McDermott}}}, \bibinfo {author} {\bibfnamefont {H.~M.}\ \bibnamefont {{van
  Horn}}},\ and\ \bibinfo {author} {\bibfnamefont {C.~J.}\ \bibnamefont
  {{Hansen}}},\ }\bibfield  {title} {\bibinfo {title} {{Nonradial Oscillations
  of Neutron Stars}},\ }\href {https://doi.org/10.1086/166044} {\bibfield
  {journal} {\bibinfo  {journal} {Astrophys. J.}\ }\textbf {\bibinfo {volume}
  {325}},\ \bibinfo {pages} {725} (\bibinfo {year} {1988})}\BibitemShut
  {NoStop}%
\bibitem [{\citenamefont {Ferrari}\ \emph {et~al.}(2003)\citenamefont
  {Ferrari}, \citenamefont {Miniutti},\ and\ \citenamefont
  {Pons}}]{Ferrari:2003nk}%
  \BibitemOpen
  \bibfield  {author} {\bibinfo {author} {\bibfnamefont {V.}~\bibnamefont
  {Ferrari}}, \bibinfo {author} {\bibfnamefont {G.}~\bibnamefont {Miniutti}},\
  and\ \bibinfo {author} {\bibfnamefont {J.~A.}\ \bibnamefont {Pons}},\
  }\bibfield  {title} {\bibinfo {title} {{Gravitational waves from neutron
  stars at different evolutionary stages}},\ }\href
  {https://doi.org/10.1088/0264-9381/20/17/327} {\bibfield  {journal} {\bibinfo
   {journal} {Class. Quant. Grav.}\ }\textbf {\bibinfo {volume} {20}},\
  \bibinfo {pages} {S841} (\bibinfo {year} {2003})}\BibitemShut {NoStop}%
\bibitem [{\citenamefont {Kr\"uger}\ \emph {et~al.}(2015)\citenamefont
  {Kr\"uger}, \citenamefont {Ho},\ and\ \citenamefont
  {Andersson}}]{Kruger:2014pva}%
  \BibitemOpen
  \bibfield  {author} {\bibinfo {author} {\bibfnamefont {C.~J.}\ \bibnamefont
  {Kr\"uger}}, \bibinfo {author} {\bibfnamefont {W.~C.~G.}\ \bibnamefont
  {Ho}},\ and\ \bibinfo {author} {\bibfnamefont {N.}~\bibnamefont
  {Andersson}},\ }\bibfield  {title} {\bibinfo {title} {{Seismology of
  adolescent neutron stars: Accounting for thermal effects and crust
  elasticity}},\ }\href {https://doi.org/10.1103/PhysRevD.92.063009} {\bibfield
   {journal} {\bibinfo  {journal} {Phys. Rev. D}\ }\textbf {\bibinfo {volume}
  {92}},\ \bibinfo {pages} {063009} (\bibinfo {year} {2015})},\ \Eprint
  {https://arxiv.org/abs/1402.5656} {arXiv:1402.5656 [gr-qc]} \BibitemShut
  {NoStop}%
\bibitem [{\citenamefont {{Lee}}(1995)}]{Lee:1995}%
  \BibitemOpen
  \bibfield  {author} {\bibinfo {author} {\bibfnamefont {U.}~\bibnamefont
  {{Lee}}},\ }\bibfield  {title} {\bibinfo {title} {{Nonradial oscillations of
  neutron stars with the superfluid core.}},\ }\href@noop {} {\bibfield
  {journal} {\bibinfo  {journal} {\aap}\ }\textbf {\bibinfo {volume} {303}},\
  \bibinfo {pages} {515} (\bibinfo {year} {1995})}\BibitemShut {NoStop}%
\bibitem [{\citenamefont {Gusakov}\ and\ \citenamefont
  {Kantor}(2013)}]{Gusakov:2013eoa}%
  \BibitemOpen
  \bibfield  {author} {\bibinfo {author} {\bibfnamefont {M.~E.}\ \bibnamefont
  {Gusakov}}\ and\ \bibinfo {author} {\bibfnamefont {E.~M.}\ \bibnamefont
  {Kantor}},\ }\bibfield  {title} {\bibinfo {title} {{Thermal $g$-modes and
  unexpected convection in superfluid neutron stars}},\ }\href
  {https://doi.org/10.1103/PhysRevD.88.101302} {\bibfield  {journal} {\bibinfo
  {journal} {Phys. Rev. D}\ }\textbf {\bibinfo {volume} {88}},\ \bibinfo
  {pages} {101302} (\bibinfo {year} {2013})}\BibitemShut {NoStop}%
\bibitem [{\citenamefont {Kantor}\ and\ \citenamefont
  {Gusakov}(2014)}]{Kantor:2014lja}%
  \BibitemOpen
  \bibfield  {author} {\bibinfo {author} {\bibfnamefont {E.~M.}\ \bibnamefont
  {Kantor}}\ and\ \bibinfo {author} {\bibfnamefont {M.~E.}\ \bibnamefont
  {Gusakov}},\ }\bibfield  {title} {\bibinfo {title} {{Composition
  temperature-dependent g-modes in superfluid neutron stars}},\ }\href
  {https://doi.org/10.1093/mnrasl/slu061} {\bibfield  {journal} {\bibinfo
  {journal} {Mon. Not. Roy. Astron. Soc.}\ }\textbf {\bibinfo {volume} {442}},\
  \bibinfo {pages} {90} (\bibinfo {year} {2014})},\ \Eprint
  {https://arxiv.org/abs/1404.6768} {arXiv:1404.6768 [astro-ph.SR]}
  \BibitemShut {NoStop}%
\bibitem [{\citenamefont {Andersson}\ and\ \citenamefont
  {Comer}(2001)}]{Andersson:2001bz}%
  \BibitemOpen
  \bibfield  {author} {\bibinfo {author} {\bibfnamefont {N.}~\bibnamefont
  {Andersson}}\ and\ \bibinfo {author} {\bibfnamefont {G.~L.}\ \bibnamefont
  {Comer}},\ }\bibfield  {title} {\bibinfo {title} {{On the dynamics of
  superfluid neutron star cores}},\ }\href
  {https://doi.org/10.1046/j.1365-8711.2001.04923.x} {\bibfield  {journal}
  {\bibinfo  {journal} {Mon. Not. Roy. Astron. Soc.}\ }\textbf {\bibinfo
  {volume} {328}},\ \bibinfo {pages} {1129} (\bibinfo {year} {2001})},\ \Eprint
  {https://arxiv.org/abs/astro-ph/0101193} {arXiv:astro-ph/0101193}
  \BibitemShut {NoStop}%
\bibitem [{\citenamefont {Passamonti}\ \emph {et~al.}(2016)\citenamefont
  {Passamonti}, \citenamefont {Andersson},\ and\ \citenamefont
  {Ho}}]{Passamonti:2015oia}%
  \BibitemOpen
  \bibfield  {author} {\bibinfo {author} {\bibfnamefont {A.}~\bibnamefont
  {Passamonti}}, \bibinfo {author} {\bibfnamefont {N.}~\bibnamefont
  {Andersson}},\ and\ \bibinfo {author} {\bibfnamefont {W.~C.~G.}\ \bibnamefont
  {Ho}},\ }\bibfield  {title} {\bibinfo {title} {{Buoyancy and g-modes in young
  superfluid neutron stars}},\ }\href {https://doi.org/10.1093/mnras/stv2149}
  {\bibfield  {journal} {\bibinfo  {journal} {Mon. Not. Roy. Astron. Soc.}\
  }\textbf {\bibinfo {volume} {455}},\ \bibinfo {pages} {1489} (\bibinfo {year}
  {2016})},\ \Eprint {https://arxiv.org/abs/1504.07470} {arXiv:1504.07470
  [astro-ph.SR]} \BibitemShut {NoStop}%
\bibitem [{\citenamefont {{Finn}}(1987)}]{Finn:1987}%
  \BibitemOpen
  \bibfield  {author} {\bibinfo {author} {\bibfnamefont {L.~S.}\ \bibnamefont
  {{Finn}}},\ }\bibfield  {title} {\bibinfo {title} {{G-modes in
  zero-temperature neutron stars}},\ }\href
  {https://doi.org/10.1093/mnras/227.2.265} {\bibfield  {journal} {\bibinfo
  {journal} {Mon. Not. Roy. Astron. Soc.}\ }\textbf {\bibinfo {volume} {227}},\
  \bibinfo {pages} {265} (\bibinfo {year} {1987})}\BibitemShut {NoStop}%
\bibitem [{\citenamefont {{McDermott}}(1990)}]{McDermott:1990}%
  \BibitemOpen
  \bibfield  {author} {\bibinfo {author} {\bibfnamefont {P.~N.}\ \bibnamefont
  {{McDermott}}},\ }\bibfield  {title} {\bibinfo {title} {{Density
  Discontinuity G-Modes}},\ }\href@noop {} {\bibfield  {journal} {\bibinfo
  {journal} {Mon. Not. Roy. Astron. Soc.}\ }\textbf {\bibinfo {volume} {245}},\
  \bibinfo {pages} {508} (\bibinfo {year} {1990})}\BibitemShut {NoStop}%
\bibitem [{\citenamefont {Sotani}\ \emph {et~al.}(2001)\citenamefont {Sotani},
  \citenamefont {Tominaga},\ and\ \citenamefont {Maeda}}]{Sotani:2001bb}%
  \BibitemOpen
  \bibfield  {author} {\bibinfo {author} {\bibfnamefont {H.}~\bibnamefont
  {Sotani}}, \bibinfo {author} {\bibfnamefont {K.}~\bibnamefont {Tominaga}},\
  and\ \bibinfo {author} {\bibfnamefont {K.-i.}\ \bibnamefont {Maeda}},\
  }\bibfield  {title} {\bibinfo {title} {{Density discontinuity of a neutron
  star and gravitational waves}},\ }\href
  {https://doi.org/10.1103/PhysRevD.65.024010} {\bibfield  {journal} {\bibinfo
  {journal} {Phys. Rev. D}\ }\textbf {\bibinfo {volume} {65}},\ \bibinfo
  {pages} {024010} (\bibinfo {year} {2001})},\ \Eprint
  {https://arxiv.org/abs/gr-qc/0108060} {arXiv:gr-qc/0108060} \BibitemShut
  {NoStop}%
\bibitem [{\citenamefont {Miniutti}\ \emph {et~al.}(2003)\citenamefont
  {Miniutti}, \citenamefont {Pons}, \citenamefont {Berti}, \citenamefont
  {Gualtieri},\ and\ \citenamefont {Ferrari}}]{Miniutti:2002bh}%
  \BibitemOpen
  \bibfield  {author} {\bibinfo {author} {\bibfnamefont {G.}~\bibnamefont
  {Miniutti}}, \bibinfo {author} {\bibfnamefont {J.~A.}\ \bibnamefont {Pons}},
  \bibinfo {author} {\bibfnamefont {E.}~\bibnamefont {Berti}}, \bibinfo
  {author} {\bibfnamefont {L.}~\bibnamefont {Gualtieri}},\ and\ \bibinfo
  {author} {\bibfnamefont {V.}~\bibnamefont {Ferrari}},\ }\bibfield  {title}
  {\bibinfo {title} {{Non-radial oscillation modes as a probe of density
  discontinuities in neutron stars}},\ }\href
  {https://doi.org/10.1046/j.1365-8711.2003.06057.x} {\bibfield  {journal}
  {\bibinfo  {journal} {Mon. Not. Roy. Astron. Soc.}\ }\textbf {\bibinfo
  {volume} {338}},\ \bibinfo {pages} {389} (\bibinfo {year} {2003})},\ \Eprint
  {https://arxiv.org/abs/astro-ph/0206142} {arXiv:astro-ph/0206142}
  \BibitemShut {NoStop}%
\bibitem [{\citenamefont {Tonetto}\ and\ \citenamefont
  {Lugones}(2020)}]{Tonetto:2020bie}%
  \BibitemOpen
  \bibfield  {author} {\bibinfo {author} {\bibfnamefont {L.}~\bibnamefont
  {Tonetto}}\ and\ \bibinfo {author} {\bibfnamefont {G.}~\bibnamefont
  {Lugones}},\ }\bibfield  {title} {\bibinfo {title} {{Discontinuity gravity
  modes in hybrid stars: assessing the role of rapid and slow phase
  conversions}},\ }\href {https://doi.org/10.1103/PhysRevD.101.123029}
  {\bibfield  {journal} {\bibinfo  {journal} {Phys. Rev. D}\ }\textbf {\bibinfo
  {volume} {101}},\ \bibinfo {pages} {123029} (\bibinfo {year} {2020})},\
  \Eprint {https://arxiv.org/abs/2003.01259} {arXiv:2003.01259 [astro-ph.HE]}
  \BibitemShut {NoStop}%
\bibitem [{\citenamefont {Constantinou}\ \emph {et~al.}(2021)\citenamefont
  {Constantinou}, \citenamefont {Han}, \citenamefont {Jaikumar},\ and\
  \citenamefont {Prakash}}]{Constantinou:2021hba}%
  \BibitemOpen
  \bibfield  {author} {\bibinfo {author} {\bibfnamefont {C.}~\bibnamefont
  {Constantinou}}, \bibinfo {author} {\bibfnamefont {S.}~\bibnamefont {Han}},
  \bibinfo {author} {\bibfnamefont {P.}~\bibnamefont {Jaikumar}},\ and\
  \bibinfo {author} {\bibfnamefont {M.}~\bibnamefont {Prakash}},\ }\bibfield
  {title} {\bibinfo {title} {{g modes of neutron stars with hadron-to-quark
  crossover transitions}},\ }\href
  {https://doi.org/10.1103/PhysRevD.104.123032} {\bibfield  {journal} {\bibinfo
   {journal} {Phys. Rev. D}\ }\textbf {\bibinfo {volume} {104}},\ \bibinfo
  {pages} {123032} (\bibinfo {year} {2021})},\ \Eprint
  {https://arxiv.org/abs/2109.14091} {arXiv:2109.14091 [astro-ph.HE]}
  \BibitemShut {NoStop}%
\bibitem [{\citenamefont {Zhao}\ \emph {et~al.}(2022)\citenamefont {Zhao},
  \citenamefont {Constantinou}, \citenamefont {Jaikumar},\ and\ \citenamefont
  {Prakash}}]{Zhao:2022toc}%
  \BibitemOpen
  \bibfield  {author} {\bibinfo {author} {\bibfnamefont {T.}~\bibnamefont
  {Zhao}}, \bibinfo {author} {\bibfnamefont {C.}~\bibnamefont {Constantinou}},
  \bibinfo {author} {\bibfnamefont {P.}~\bibnamefont {Jaikumar}},\ and\
  \bibinfo {author} {\bibfnamefont {M.}~\bibnamefont {Prakash}},\ }\bibfield
  {title} {\bibinfo {title} {{Quasinormal g modes of neutron stars with
  quarks}},\ }\href {https://doi.org/10.1103/PhysRevD.105.103025} {\bibfield
  {journal} {\bibinfo  {journal} {Phys. Rev. D}\ }\textbf {\bibinfo {volume}
  {105}},\ \bibinfo {pages} {103025} (\bibinfo {year} {2022})},\ \Eprint
  {https://arxiv.org/abs/2202.01403} {arXiv:2202.01403 [gr-qc]} \BibitemShut
  {NoStop}%
\bibitem [{\citenamefont {Abbott}\ \emph {et~al.}(2017)\citenamefont {Abbott}
  \emph {et~al.}}]{LIGOScientific:2017vwq}%
  \BibitemOpen
  \bibfield  {author} {\bibinfo {author} {\bibfnamefont {B.~P.}\ \bibnamefont
  {Abbott}} \emph {et~al.} (\bibinfo {collaboration} {LIGO Scientific,
  Virgo}),\ }\bibfield  {title} {\bibinfo {title} {{GW170817: Observation of
  Gravitational Waves from a Binary Neutron Star Inspiral}},\ }\href
  {https://doi.org/10.1103/PhysRevLett.119.161101} {\bibfield  {journal}
  {\bibinfo  {journal} {Phys. Rev. Lett.}\ }\textbf {\bibinfo {volume} {119}},\
  \bibinfo {pages} {161101} (\bibinfo {year} {2017})},\ \Eprint
  {https://arxiv.org/abs/1710.05832} {arXiv:1710.05832 [gr-qc]} \BibitemShut
  {NoStop}%
\bibitem [{\citenamefont {Abbott}\ \emph {et~al.}(2018)\citenamefont {Abbott}
  \emph {et~al.}}]{LIGOScientific:2018cki}%
  \BibitemOpen
  \bibfield  {author} {\bibinfo {author} {\bibfnamefont {B.~P.}\ \bibnamefont
  {Abbott}} \emph {et~al.} (\bibinfo {collaboration} {LIGO Scientific,
  Virgo}),\ }\bibfield  {title} {\bibinfo {title} {{GW170817: Measurements of
  neutron star radii and equation of state}},\ }\href
  {https://doi.org/10.1103/PhysRevLett.121.161101} {\bibfield  {journal}
  {\bibinfo  {journal} {Phys. Rev. Lett.}\ }\textbf {\bibinfo {volume} {121}},\
  \bibinfo {pages} {161101} (\bibinfo {year} {2018})},\ \Eprint
  {https://arxiv.org/abs/1805.11581} {arXiv:1805.11581 [gr-qc]} \BibitemShut
  {NoStop}%
\bibitem [{\citenamefont {Li}\ \emph {et~al.}(2022)\citenamefont {Li},
  \citenamefont {Gao}, \citenamefont {Shao}, \citenamefont {Xu},\ and\
  \citenamefont {Xu}}]{Li:2022qql}%
  \BibitemOpen
  \bibfield  {author} {\bibinfo {author} {\bibfnamefont {H.-B.}\ \bibnamefont
  {Li}}, \bibinfo {author} {\bibfnamefont {Y.}~\bibnamefont {Gao}}, \bibinfo
  {author} {\bibfnamefont {L.}~\bibnamefont {Shao}}, \bibinfo {author}
  {\bibfnamefont {R.-X.}\ \bibnamefont {Xu}},\ and\ \bibinfo {author}
  {\bibfnamefont {R.}~\bibnamefont {Xu}},\ }\bibfield  {title} {\bibinfo
  {title} {{Oscillation modes and gravitational waves from strangeon stars}},\
  }\href {https://doi.org/10.1093/mnras/stac2622} {\bibfield  {journal}
  {\bibinfo  {journal} {Mon. Not. Roy. Astron. Soc.}\ }\textbf {\bibinfo
  {volume} {516}},\ \bibinfo {pages} {6172} (\bibinfo {year} {2022})},\ \Eprint
  {https://arxiv.org/abs/2206.09407} {arXiv:2206.09407 [gr-qc]} \BibitemShut
  {NoStop}%
\bibitem [{\citenamefont {Lai}(1994)}]{Lai:1993di}%
  \BibitemOpen
  \bibfield  {author} {\bibinfo {author} {\bibfnamefont {D.}~\bibnamefont
  {Lai}},\ }\bibfield  {title} {\bibinfo {title} {{Resonant oscillations and
  tidal heating in coalescing binary neutron stars}},\ }\href
  {https://doi.org/10.1093/mnras/270.3.611} {\bibfield  {journal} {\bibinfo
  {journal} {Mon. Not. Roy. Astron. Soc.}\ }\textbf {\bibinfo {volume} {270}},\
  \bibinfo {pages} {611} (\bibinfo {year} {1994})},\ \Eprint
  {https://arxiv.org/abs/astro-ph/9404062} {arXiv:astro-ph/9404062}
  \BibitemShut {NoStop}%
\bibitem [{\citenamefont {Kuan}\ \emph {et~al.}(2021)\citenamefont {Kuan},
  \citenamefont {Suvorov},\ and\ \citenamefont {Kokkotas}}]{Kuan:2021jmk}%
  \BibitemOpen
  \bibfield  {author} {\bibinfo {author} {\bibfnamefont {H.-J.}\ \bibnamefont
  {Kuan}}, \bibinfo {author} {\bibfnamefont {A.~G.}\ \bibnamefont {Suvorov}},\
  and\ \bibinfo {author} {\bibfnamefont {K.~D.}\ \bibnamefont {Kokkotas}},\
  }\bibfield  {title} {\bibinfo {title} {{General-relativistic treatment of
  tidal g-mode resonances in coalescing binaries of neutron stars \textendash{}
  I. Theoretical framework and crust breaking}},\ }\href
  {https://doi.org/10.1093/mnras/stab1898} {\bibfield  {journal} {\bibinfo
  {journal} {Mon. Not. Roy. Astron. Soc.}\ }\textbf {\bibinfo {volume} {506}},\
  \bibinfo {pages} {2985} (\bibinfo {year} {2021})},\ \Eprint
  {https://arxiv.org/abs/2106.16123} {arXiv:2106.16123 [gr-qc]} \BibitemShut
  {NoStop}%
\bibitem [{\citenamefont {Lai}\ and\ \citenamefont {Wu}(2006)}]{Lai:2006pr}%
  \BibitemOpen
  \bibfield  {author} {\bibinfo {author} {\bibfnamefont {D.}~\bibnamefont
  {Lai}}\ and\ \bibinfo {author} {\bibfnamefont {Y.}~\bibnamefont {Wu}},\
  }\bibfield  {title} {\bibinfo {title} {{Resonant Tidal Excitations of
  Inertial Modes in Coalescing Neutron Star Binaries}},\ }\href
  {https://doi.org/10.1103/PhysRevD.74.024007} {\bibfield  {journal} {\bibinfo
  {journal} {Phys. Rev. D}\ }\textbf {\bibinfo {volume} {74}},\ \bibinfo
  {pages} {024007} (\bibinfo {year} {2006})},\ \Eprint
  {https://arxiv.org/abs/astro-ph/0604163} {arXiv:astro-ph/0604163}
  \BibitemShut {NoStop}%
\bibitem [{\citenamefont {Xu}\ and\ \citenamefont {Lai}(2017)}]{Xu:2017hqo}%
  \BibitemOpen
  \bibfield  {author} {\bibinfo {author} {\bibfnamefont {W.}~\bibnamefont
  {Xu}}\ and\ \bibinfo {author} {\bibfnamefont {D.}~\bibnamefont {Lai}},\
  }\bibfield  {title} {\bibinfo {title} {{Resonant Tidal Excitation of
  Oscillation Modes in Merging Binary Neutron Stars: Inertial-Gravity Modes}},\
  }\href {https://doi.org/10.1103/PhysRevD.96.083005} {\bibfield  {journal}
  {\bibinfo  {journal} {Phys. Rev. D}\ }\textbf {\bibinfo {volume} {96}},\
  \bibinfo {pages} {083005} (\bibinfo {year} {2017})},\ \Eprint
  {https://arxiv.org/abs/1708.01839} {arXiv:1708.01839 [astro-ph.HE]}
  \BibitemShut {NoStop}%
\bibitem [{\citenamefont {Lai}(1999)}]{Lai:1998yc}%
  \BibitemOpen
  \bibfield  {author} {\bibinfo {author} {\bibfnamefont {D.}~\bibnamefont
  {Lai}},\ }\bibfield  {title} {\bibinfo {title} {{Secular instability of g
  modes in rotating neutron stars}},\ }\href
  {https://doi.org/10.1046/j.1365-8711.1999.02723.x} {\bibfield  {journal}
  {\bibinfo  {journal} {Mon. Not. Roy. Astron. Soc.}\ }\textbf {\bibinfo
  {volume} {307}},\ \bibinfo {pages} {1001} (\bibinfo {year} {1999})},\ \Eprint
  {https://arxiv.org/abs/astro-ph/9806378} {arXiv:astro-ph/9806378}
  \BibitemShut {NoStop}%
\bibitem [{\citenamefont {Gaertig}\ and\ \citenamefont
  {Kokkotas}(2009)}]{Gaertig:2009rr}%
  \BibitemOpen
  \bibfield  {author} {\bibinfo {author} {\bibfnamefont {E.}~\bibnamefont
  {Gaertig}}\ and\ \bibinfo {author} {\bibfnamefont {K.~D.}\ \bibnamefont
  {Kokkotas}},\ }\bibfield  {title} {\bibinfo {title} {{Relativistic g-modes in
  rapidly rotating neutron stars}},\ }\href
  {https://doi.org/10.1103/PhysRevD.80.064026} {\bibfield  {journal} {\bibinfo
  {journal} {Phys. Rev. D}\ }\textbf {\bibinfo {volume} {80}},\ \bibinfo
  {pages} {064026} (\bibinfo {year} {2009})},\ \Eprint
  {https://arxiv.org/abs/0905.0821} {arXiv:0905.0821 [astro-ph.SR]}
  \BibitemShut {NoStop}%
\bibitem [{\citenamefont {Weinberg}\ \emph {et~al.}(2013)\citenamefont
  {Weinberg}, \citenamefont {Arras},\ and\ \citenamefont
  {Burkart}}]{Weinberg:2013pbi}%
  \BibitemOpen
  \bibfield  {author} {\bibinfo {author} {\bibfnamefont {N.~N.}\ \bibnamefont
  {Weinberg}}, \bibinfo {author} {\bibfnamefont {P.}~\bibnamefont {Arras}},\
  and\ \bibinfo {author} {\bibfnamefont {J.}~\bibnamefont {Burkart}},\
  }\bibfield  {title} {\bibinfo {title} {{An instability due to the nonlinear
  coupling of p-modes to g-modes: Implications for coalescing neutron star
  binaries}},\ }\href {https://doi.org/10.1088/0004-637X/769/2/121} {\bibfield
  {journal} {\bibinfo  {journal} {Astrophys. J.}\ }\textbf {\bibinfo {volume}
  {769}},\ \bibinfo {pages} {121} (\bibinfo {year} {2013})},\ \Eprint
  {https://arxiv.org/abs/1302.2292} {arXiv:1302.2292 [astro-ph.SR]}
  \BibitemShut {NoStop}%
\bibitem [{\citenamefont {Abbott}\ \emph {et~al.}(2019)\citenamefont {Abbott}
  \emph {et~al.}}]{LIGOScientific:2018ehx}%
  \BibitemOpen
  \bibfield  {author} {\bibinfo {author} {\bibfnamefont {B.~P.}\ \bibnamefont
  {Abbott}} \emph {et~al.} (\bibinfo {collaboration} {LIGO Scientific,
  Virgo}),\ }\bibfield  {title} {\bibinfo {title} {{Constraining the
  $p$-Mode\textendash{}$g$-Mode Tidal Instability with GW170817}},\ }\href
  {https://doi.org/10.1103/PhysRevLett.122.061104} {\bibfield  {journal}
  {\bibinfo  {journal} {Phys. Rev. Lett.}\ }\textbf {\bibinfo {volume} {122}},\
  \bibinfo {pages} {061104} (\bibinfo {year} {2019})},\ \Eprint
  {https://arxiv.org/abs/1808.08676} {arXiv:1808.08676 [astro-ph.HE]}
  \BibitemShut {NoStop}%
\bibitem [{\citenamefont {Kuan}\ \emph {et~al.}(2022)\citenamefont {Kuan},
  \citenamefont {Kr\"uger}, \citenamefont {Suvorov},\ and\ \citenamefont
  {Kokkotas}}]{Kuan:2022bhu}%
  \BibitemOpen
  \bibfield  {author} {\bibinfo {author} {\bibfnamefont {H.-J.}\ \bibnamefont
  {Kuan}}, \bibinfo {author} {\bibfnamefont {C.~J.}\ \bibnamefont {Kr\"uger}},
  \bibinfo {author} {\bibfnamefont {A.~G.}\ \bibnamefont {Suvorov}},\ and\
  \bibinfo {author} {\bibfnamefont {K.~D.}\ \bibnamefont {Kokkotas}},\
  }\bibfield  {title} {\bibinfo {title} {{Constraining equation-of-state groups
  from g-mode asteroseismology}},\ }\href
  {https://doi.org/10.1093/mnras/stac1101} {\bibfield  {journal} {\bibinfo
  {journal} {Mon. Not. Roy. Astron. Soc.}\ }\textbf {\bibinfo {volume} {513}},\
  \bibinfo {pages} {4045} (\bibinfo {year} {2022})},\ \Eprint
  {https://arxiv.org/abs/2204.08492} {arXiv:2204.08492 [gr-qc]} \BibitemShut
  {NoStop}%
\bibitem [{\citenamefont {Andersson}\ and\ \citenamefont
  {Pnigouras}(2019)}]{Andersson:2019mxp}%
  \BibitemOpen
  \bibfield  {author} {\bibinfo {author} {\bibfnamefont {N.}~\bibnamefont
  {Andersson}}\ and\ \bibinfo {author} {\bibfnamefont {P.}~\bibnamefont
  {Pnigouras}},\ }\bibfield  {title} {\bibinfo {title} {{The g-mode spectrum of
  reactive neutron star cores}},\ }\href
  {https://doi.org/10.1093/mnras/stz2449} {\bibfield  {journal} {\bibinfo
  {journal} {Mon. Not. Roy. Astron. Soc.}\ }\textbf {\bibinfo {volume} {489}},\
  \bibinfo {pages} {4043} (\bibinfo {year} {2019})},\ \Eprint
  {https://arxiv.org/abs/1905.00010} {arXiv:1905.00010 [gr-qc]} \BibitemShut
  {NoStop}%
\bibitem [{\citenamefont {Fu}\ \emph {et~al.}(2008)\citenamefont {Fu},
  \citenamefont {Wei},\ and\ \citenamefont {Liu}}]{Fu:2008bu}%
  \BibitemOpen
  \bibfield  {author} {\bibinfo {author} {\bibfnamefont {W.-J.}\ \bibnamefont
  {Fu}}, \bibinfo {author} {\bibfnamefont {H.-Q.}\ \bibnamefont {Wei}},\ and\
  \bibinfo {author} {\bibfnamefont {Y.-X.}\ \bibnamefont {Liu}},\ }\bibfield
  {title} {\bibinfo {title} {{Distinguishing Newly Born Strange Stars from
  Neutron Stars with g-Mode Oscillations}},\ }\href
  {https://doi.org/10.1103/PhysRevLett.101.181102} {\bibfield  {journal}
  {\bibinfo  {journal} {Phys. Rev. Lett.}\ }\textbf {\bibinfo {volume} {101}},\
  \bibinfo {pages} {181102} (\bibinfo {year} {2008})},\ \Eprint
  {https://arxiv.org/abs/0810.1084} {arXiv:0810.1084 [nucl-th]} \BibitemShut
  {NoStop}%
\bibitem [{\citenamefont {Ott}\ \emph {et~al.}(2006)\citenamefont {Ott},
  \citenamefont {Burrows}, \citenamefont {Dessart},\ and\ \citenamefont
  {Livne}}]{Ott:2006qp}%
  \BibitemOpen
  \bibfield  {author} {\bibinfo {author} {\bibfnamefont {C.~D.}\ \bibnamefont
  {Ott}}, \bibinfo {author} {\bibfnamefont {A.}~\bibnamefont {Burrows}},
  \bibinfo {author} {\bibfnamefont {L.}~\bibnamefont {Dessart}},\ and\ \bibinfo
  {author} {\bibfnamefont {E.}~\bibnamefont {Livne}},\ }\bibfield  {title}
  {\bibinfo {title} {{A New Mechanism for Gravitational Wave Emission in
  Core-Collapse Supernovae}},\ }\href
  {https://doi.org/10.1103/PhysRevLett.96.201102} {\bibfield  {journal}
  {\bibinfo  {journal} {Phys. Rev. Lett.}\ }\textbf {\bibinfo {volume} {96}},\
  \bibinfo {pages} {201102} (\bibinfo {year} {2006})},\ \Eprint
  {https://arxiv.org/abs/astro-ph/0605493} {arXiv:astro-ph/0605493}
  \BibitemShut {NoStop}%
\bibitem [{\citenamefont {Pajkos}\ \emph {et~al.}(2019)\citenamefont {Pajkos},
  \citenamefont {Couch}, \citenamefont {Pan},\ and\ \citenamefont
  {O'Connor}}]{Pajkos:2019nef}%
  \BibitemOpen
  \bibfield  {author} {\bibinfo {author} {\bibfnamefont {M.~A.}\ \bibnamefont
  {Pajkos}}, \bibinfo {author} {\bibfnamefont {S.~M.}\ \bibnamefont {Couch}},
  \bibinfo {author} {\bibfnamefont {K.-C.}\ \bibnamefont {Pan}},\ and\ \bibinfo
  {author} {\bibfnamefont {E.~P.}\ \bibnamefont {O'Connor}},\ }\bibfield
  {title} {\bibinfo {title} {{Features of Accretion Phase Gravitational Wave
  Emission from Two-dimensional Rotating Core-Collapse Supernovae}},\ }\href
  {https://doi.org/10.3847/1538-4357/ab1de2} {\bibfield  {journal} {\bibinfo
  {journal} {Astrophys. J.}\ }\textbf {\bibinfo {volume} {878}},\ \bibinfo
  {pages} {13} (\bibinfo {year} {2019})},\ \Eprint
  {https://arxiv.org/abs/1901.09055} {arXiv:1901.09055 [astro-ph.HE]}
  \BibitemShut {NoStop}%
\bibitem [{\citenamefont
  {Westernacher-Schneider}(2020)}]{Westernacher-Schneider:2020bkw}%
  \BibitemOpen
  \bibfield  {author} {\bibinfo {author} {\bibfnamefont {J.~R.}\ \bibnamefont
  {Westernacher-Schneider}},\ }\bibfield  {title} {\bibinfo {title}
  {{Consistent perturbative modeling of pseudo-Newtonian core-collapse
  supernova simulations}},\ }\href
  {https://doi.org/10.1103/PhysRevD.101.083021} {\bibfield  {journal} {\bibinfo
   {journal} {Phys. Rev. D}\ }\textbf {\bibinfo {volume} {101}},\ \bibinfo
  {pages} {083021} (\bibinfo {year} {2020})},\ \Eprint
  {https://arxiv.org/abs/2002.04468} {arXiv:2002.04468 [astro-ph.HE]}
  \BibitemShut {NoStop}%
\bibitem [{\citenamefont {{Lagaris}}\ and\ \citenamefont
  {{Pandharipande}}(1981)}]{Lagaris:1981}%
  \BibitemOpen
  \bibfield  {author} {\bibinfo {author} {\bibfnamefont {I.~E.}\ \bibnamefont
  {{Lagaris}}}\ and\ \bibinfo {author} {\bibfnamefont {V.~R.}\ \bibnamefont
  {{Pandharipande}}},\ }\bibfield  {title} {\bibinfo {title} {{Variational
  calculations of asymmetric nuclear matter}},\ }\href
  {https://doi.org/10.1016/0375-9474(81)90032-4} {\bibfield  {journal}
  {\bibinfo  {journal} {Nucl. Phys. A}\ }\textbf {\bibinfo {volume} {369}},\
  \bibinfo {pages} {470} (\bibinfo {year} {1981})}\BibitemShut {NoStop}%
\bibitem [{\citenamefont {{Prakash}}\ \emph {et~al.}(1988)\citenamefont
  {{Prakash}}, \citenamefont {{Ainsworth}},\ and\ \citenamefont
  {{Lattimer}}}]{Prakash:1988}%
  \BibitemOpen
  \bibfield  {author} {\bibinfo {author} {\bibfnamefont {M.}~\bibnamefont
  {{Prakash}}}, \bibinfo {author} {\bibfnamefont {T.~L.}\ \bibnamefont
  {{Ainsworth}}},\ and\ \bibinfo {author} {\bibfnamefont {J.~M.}\ \bibnamefont
  {{Lattimer}}},\ }\bibfield  {title} {\bibinfo {title} {{Equation of state and
  the maximum mass of neutron stars}},\ }\href
  {https://doi.org/10.1103/PhysRevLett.61.2518} {\bibfield  {journal} {\bibinfo
   {journal} {\prl}\ }\textbf {\bibinfo {volume} {61}},\ \bibinfo {pages}
  {2518} (\bibinfo {year} {1988})}\BibitemShut {NoStop}%
\bibitem [{\citenamefont {Wiringa}\ \emph {et~al.}(1988)\citenamefont
  {Wiringa}, \citenamefont {Fiks},\ and\ \citenamefont
  {Fabrocini}}]{Wiringa:1988tp}%
  \BibitemOpen
  \bibfield  {author} {\bibinfo {author} {\bibfnamefont {R.~B.}\ \bibnamefont
  {Wiringa}}, \bibinfo {author} {\bibfnamefont {V.}~\bibnamefont {Fiks}},\ and\
  \bibinfo {author} {\bibfnamefont {A.}~\bibnamefont {Fabrocini}},\ }\bibfield
  {title} {\bibinfo {title} {{Equation of state for dense nucleon matter}},\
  }\href {https://doi.org/10.1103/PhysRevC.38.1010} {\bibfield  {journal}
  {\bibinfo  {journal} {Phys. Rev. C}\ }\textbf {\bibinfo {volume} {38}},\
  \bibinfo {pages} {1010} (\bibinfo {year} {1988})}\BibitemShut {NoStop}%
\bibitem [{\citenamefont {Lattimer}(2014)}]{Lattimer:2014scr}%
  \BibitemOpen
  \bibfield  {author} {\bibinfo {author} {\bibfnamefont {J.~M.}\ \bibnamefont
  {Lattimer}},\ }\bibfield  {title} {\bibinfo {title} {{Symmetry energy in
  nuclei and neutron stars}},\ }\href
  {https://doi.org/10.1016/j.nuclphysa.2014.04.008} {\bibfield  {journal}
  {\bibinfo  {journal} {Nucl. Phys. A}\ }\textbf {\bibinfo {volume} {928}},\
  \bibinfo {pages} {276} (\bibinfo {year} {2014})}\BibitemShut {NoStop}%
\bibitem [{\citenamefont {{Baym}}\ \emph
  {et~al.}(1971{\natexlab{a}})\citenamefont {{Baym}}, \citenamefont {{Bethe}},\
  and\ \citenamefont {{Pethick}}}]{BBP:1971}%
  \BibitemOpen
  \bibfield  {author} {\bibinfo {author} {\bibfnamefont {G.}~\bibnamefont
  {{Baym}}}, \bibinfo {author} {\bibfnamefont {H.~A.}\ \bibnamefont
  {{Bethe}}},\ and\ \bibinfo {author} {\bibfnamefont {C.~J.}\ \bibnamefont
  {{Pethick}}},\ }\bibfield  {title} {\bibinfo {title} {{Neutron star
  matter}},\ }\href {https://doi.org/10.1016/0375-9474(71)90281-8} {\bibfield
  {journal} {\bibinfo  {journal} {Nucl. Phys. A}\ }\textbf {\bibinfo {volume}
  {175}},\ \bibinfo {pages} {225} (\bibinfo {year}
  {1971}{\natexlab{a}})}\BibitemShut {NoStop}%
\bibitem [{\citenamefont {{Baym}}\ \emph
  {et~al.}(1971{\natexlab{b}})\citenamefont {{Baym}}, \citenamefont
  {{Pethick}},\ and\ \citenamefont {{Sutherland}}}]{BPS:1971}%
  \BibitemOpen
  \bibfield  {author} {\bibinfo {author} {\bibfnamefont {G.}~\bibnamefont
  {{Baym}}}, \bibinfo {author} {\bibfnamefont {C.}~\bibnamefont {{Pethick}}},\
  and\ \bibinfo {author} {\bibfnamefont {P.}~\bibnamefont {{Sutherland}}},\
  }\bibfield  {title} {\bibinfo {title} {{The Ground State of Matter at High
  Densities: Equation of State and Stellar Models}},\ }\href
  {https://doi.org/10.1086/151216} {\bibfield  {journal} {\bibinfo  {journal}
  {Astrophys. J.}\ }\textbf {\bibinfo {volume} {170}},\ \bibinfo {pages} {299}
  (\bibinfo {year} {1971}{\natexlab{b}})}\BibitemShut {NoStop}%
\bibitem [{\citenamefont {Antoniadis}\ \emph {et~al.}(2013)\citenamefont
  {Antoniadis} \emph {et~al.}}]{Antoniadis:2013pzd}%
  \BibitemOpen
  \bibfield  {author} {\bibinfo {author} {\bibfnamefont {J.}~\bibnamefont
  {Antoniadis}} \emph {et~al.},\ }\bibfield  {title} {\bibinfo {title} {{A
  Massive Pulsar in a Compact Relativistic Binary}},\ }\href
  {https://doi.org/10.1126/science.1233232} {\bibfield  {journal} {\bibinfo
  {journal} {Science}\ }\textbf {\bibinfo {volume} {340}},\ \bibinfo {pages}
  {6131} (\bibinfo {year} {2013})},\ \Eprint {https://arxiv.org/abs/1304.6875}
  {arXiv:1304.6875 [astro-ph.HE]} \BibitemShut {NoStop}%
\bibitem [{\citenamefont {Fonseca}\ \emph {et~al.}(2021)\citenamefont {Fonseca}
  \emph {et~al.}}]{Fonseca:2021wxt}%
  \BibitemOpen
  \bibfield  {author} {\bibinfo {author} {\bibfnamefont {E.}~\bibnamefont
  {Fonseca}} \emph {et~al.},\ }\bibfield  {title} {\bibinfo {title} {{Refined
  Mass and Geometric Measurements of the High-mass PSR J0740+6620}},\ }\href
  {https://doi.org/10.3847/2041-8213/ac03b8} {\bibfield  {journal} {\bibinfo
  {journal} {Astrophys. J. Lett.}\ }\textbf {\bibinfo {volume} {915}},\
  \bibinfo {pages} {L12} (\bibinfo {year} {2021})},\ \Eprint
  {https://arxiv.org/abs/2104.00880} {arXiv:2104.00880 [astro-ph.HE]}
  \BibitemShut {NoStop}%
\bibitem [{\citenamefont {{Lindblom}}\ and\ \citenamefont
  {{Detweiler}}(1983)}]{Lindblom:1983}%
  \BibitemOpen
  \bibfield  {author} {\bibinfo {author} {\bibfnamefont {L.}~\bibnamefont
  {{Lindblom}}}\ and\ \bibinfo {author} {\bibfnamefont {S.~L.}\ \bibnamefont
  {{Detweiler}}},\ }\bibfield  {title} {\bibinfo {title} {{The quadrupole
  oscillations of neutron stars.}},\ }\href {https://doi.org/10.1086/190884}
  {\bibfield  {journal} {\bibinfo  {journal} {Astrophys. J. Suppl.}\ }\textbf
  {\bibinfo {volume} {53}},\ \bibinfo {pages} {73} (\bibinfo {year}
  {1983})}\BibitemShut {NoStop}%
\bibitem [{\citenamefont {{Detweiler}}\ and\ \citenamefont
  {{Lindblom}}(1985)}]{Detweiler:1985}%
  \BibitemOpen
  \bibfield  {author} {\bibinfo {author} {\bibfnamefont {S.}~\bibnamefont
  {{Detweiler}}}\ and\ \bibinfo {author} {\bibfnamefont {L.}~\bibnamefont
  {{Lindblom}}},\ }\bibfield  {title} {\bibinfo {title} {{On the nonradial
  pulsations of general relativistic stellar models}},\ }\href
  {https://doi.org/10.1086/163127} {\bibfield  {journal} {\bibinfo  {journal}
  {Astrophys. J.}\ }\textbf {\bibinfo {volume} {292}},\ \bibinfo {pages} {12}
  (\bibinfo {year} {1985})}\BibitemShut {NoStop}%
\bibitem [{\citenamefont {Bernuzzi}\ \emph {et~al.}(2015)\citenamefont
  {Bernuzzi}, \citenamefont {Dietrich},\ and\ \citenamefont
  {Nagar}}]{Bernuzzi:2015rla}%
  \BibitemOpen
  \bibfield  {author} {\bibinfo {author} {\bibfnamefont {S.}~\bibnamefont
  {Bernuzzi}}, \bibinfo {author} {\bibfnamefont {T.}~\bibnamefont {Dietrich}},\
  and\ \bibinfo {author} {\bibfnamefont {A.}~\bibnamefont {Nagar}},\ }\bibfield
   {title} {\bibinfo {title} {{Modeling the complete gravitational wave
  spectrum of neutron star mergers}},\ }\href
  {https://doi.org/10.1103/PhysRevLett.115.091101} {\bibfield  {journal}
  {\bibinfo  {journal} {Phys. Rev. Lett.}\ }\textbf {\bibinfo {volume} {115}},\
  \bibinfo {pages} {091101} (\bibinfo {year} {2015})},\ \Eprint
  {https://arxiv.org/abs/1504.01764} {arXiv:1504.01764 [gr-qc]} \BibitemShut
  {NoStop}%
\bibitem [{\citenamefont {Chakravarti}\ and\ \citenamefont
  {Andersson}(2020)}]{Chakravarti:2019sdc}%
  \BibitemOpen
  \bibfield  {author} {\bibinfo {author} {\bibfnamefont {K.}~\bibnamefont
  {Chakravarti}}\ and\ \bibinfo {author} {\bibfnamefont {N.}~\bibnamefont
  {Andersson}},\ }\bibfield  {title} {\bibinfo {title} {{Exploring universality
  in neutron star mergers}},\ }\href {https://doi.org/10.1093/mnras/staa2342}
  {\bibfield  {journal} {\bibinfo  {journal} {Mon. Not. Roy. Astron. Soc.}\
  }\textbf {\bibinfo {volume} {497}},\ \bibinfo {pages} {5480} (\bibinfo {year}
  {2020})},\ \Eprint {https://arxiv.org/abs/1906.04546} {arXiv:1906.04546
  [gr-qc]} \BibitemShut {NoStop}%
\bibitem [{\citenamefont {Andersson}\ \emph {et~al.}(2023)\citenamefont
  {Andersson}, \citenamefont {Gittins}, \citenamefont {Yin},\ and\
  \citenamefont {Panosso~Macedo}}]{Andersson:2022cax}%
  \BibitemOpen
  \bibfield  {author} {\bibinfo {author} {\bibfnamefont {N.}~\bibnamefont
  {Andersson}}, \bibinfo {author} {\bibfnamefont {F.}~\bibnamefont {Gittins}},
  \bibinfo {author} {\bibfnamefont {S.}~\bibnamefont {Yin}},\ and\ \bibinfo
  {author} {\bibfnamefont {R.}~\bibnamefont {Panosso~Macedo}},\ }\bibfield
  {title} {\bibinfo {title} {{Building post-Newtonian neutron stars}},\ }\href
  {https://doi.org/10.1088/1361-6382/acace5} {\bibfield  {journal} {\bibinfo
  {journal} {Class. Quant. Grav.}\ }\textbf {\bibinfo {volume} {40}},\ \bibinfo
  {pages} {025016} (\bibinfo {year} {2023})},\ \Eprint
  {https://arxiv.org/abs/2209.05871} {arXiv:2209.05871 [gr-qc]} \BibitemShut
  {NoStop}%
\bibitem [{\citenamefont {Orsaria}\ \emph {et~al.}(2019)\citenamefont
  {Orsaria}, \citenamefont {Malfatti}, \citenamefont {Mariani}, \citenamefont
  {Ranea-Sandoval}, \citenamefont {Garc\'\i{}a}, \citenamefont {Spinella},
  \citenamefont {Contrera}, \citenamefont {Lugones},\ and\ \citenamefont
  {Weber}}]{Orsaria:2019ftf}%
  \BibitemOpen
  \bibfield  {author} {\bibinfo {author} {\bibfnamefont {M.~G.}\ \bibnamefont
  {Orsaria}}, \bibinfo {author} {\bibfnamefont {G.}~\bibnamefont {Malfatti}},
  \bibinfo {author} {\bibfnamefont {M.}~\bibnamefont {Mariani}}, \bibinfo
  {author} {\bibfnamefont {I.~F.}\ \bibnamefont {Ranea-Sandoval}}, \bibinfo
  {author} {\bibfnamefont {F.}~\bibnamefont {Garc\'\i{}a}}, \bibinfo {author}
  {\bibfnamefont {W.~M.}\ \bibnamefont {Spinella}}, \bibinfo {author}
  {\bibfnamefont {G.~A.}\ \bibnamefont {Contrera}}, \bibinfo {author}
  {\bibfnamefont {G.}~\bibnamefont {Lugones}},\ and\ \bibinfo {author}
  {\bibfnamefont {F.}~\bibnamefont {Weber}},\ }\bibfield  {title} {\bibinfo
  {title} {{Phase transitions in neutron stars and their links to gravitational
  waves}},\ }\href {https://doi.org/10.1088/1361-6471/ab1d81} {\bibfield
  {journal} {\bibinfo  {journal} {J. Phys. G}\ }\textbf {\bibinfo {volume}
  {46}},\ \bibinfo {pages} {073002} (\bibinfo {year} {2019})},\ \Eprint
  {https://arxiv.org/abs/1907.04654} {arXiv:1907.04654 [astro-ph.HE]}
  \BibitemShut {NoStop}%
\bibitem [{\citenamefont {{Cutler}}\ \emph {et~al.}(1990)\citenamefont
  {{Cutler}}, \citenamefont {{Lindblom}},\ and\ \citenamefont
  {{Splinter}}}]{Cutler:1990}%
  \BibitemOpen
  \bibfield  {author} {\bibinfo {author} {\bibfnamefont {C.}~\bibnamefont
  {{Cutler}}}, \bibinfo {author} {\bibfnamefont {L.}~\bibnamefont
  {{Lindblom}}},\ and\ \bibinfo {author} {\bibfnamefont {R.~J.}\ \bibnamefont
  {{Splinter}}},\ }\bibfield  {title} {\bibinfo {title} {{Damping Times for
  Neutron Star Oscillations}},\ }\href {https://doi.org/10.1086/169370}
  {\bibfield  {journal} {\bibinfo  {journal} {Astrophys. J.}\ }\textbf
  {\bibinfo {volume} {363}},\ \bibinfo {pages} {603} (\bibinfo {year}
  {1990})}\BibitemShut {NoStop}%
\bibitem [{\citenamefont {Chugunov}\ and\ \citenamefont
  {Gusakov}(2011)}]{Chugunov:2011dc}%
  \BibitemOpen
  \bibfield  {author} {\bibinfo {author} {\bibfnamefont {A.~I.}\ \bibnamefont
  {Chugunov}}\ and\ \bibinfo {author} {\bibfnamefont {M.~E.}\ \bibnamefont
  {Gusakov}},\ }\bibfield  {title} {\bibinfo {title} {{Nonradial superfluid
  modes in oscillating neutron stars}},\ }\href
  {https://doi.org/10.1111/j.1745-3933.2011.01142.x} {\bibfield  {journal}
  {\bibinfo  {journal} {Mon. Not. Roy. Astron. Soc.}\ }\textbf {\bibinfo
  {volume} {418}},\ \bibinfo {pages} {L54} (\bibinfo {year} {2011})},\ \Eprint
  {https://arxiv.org/abs/1107.4242} {arXiv:1107.4242 [astro-ph.SR]}
  \BibitemShut {NoStop}%
\bibitem [{\citenamefont {{Gusakov}}\ and\ \citenamefont
  {{Kantor}}(2011)}]{Gusakov:2011}%
  \BibitemOpen
  \bibfield  {author} {\bibinfo {author} {\bibfnamefont {M.~E.}\ \bibnamefont
  {{Gusakov}}}\ and\ \bibinfo {author} {\bibfnamefont {E.~M.}\ \bibnamefont
  {{Kantor}}},\ }\bibfield  {title} {\bibinfo {title} {{Decoupling of
  superfluid and normal modes in pulsating neutron stars}},\ }\href
  {https://doi.org/10.1103/PhysRevD.83.081304} {\bibfield  {journal} {\bibinfo
  {journal} {\prd}\ }\textbf {\bibinfo {volume} {83}},\ \bibinfo {eid} {081304}
  (\bibinfo {year} {2011})},\ \Eprint {https://arxiv.org/abs/1007.2752}
  {arXiv:1007.2752 [astro-ph.SR]} \BibitemShut {NoStop}%
\bibitem [{\citenamefont {Gusakov}\ \emph {et~al.}(2013)\citenamefont
  {Gusakov}, \citenamefont {Kantor}, \citenamefont {Chugunov},\ and\
  \citenamefont {Gualtieri}}]{Gusakov:2012zx}%
  \BibitemOpen
  \bibfield  {author} {\bibinfo {author} {\bibfnamefont {M.~E.}\ \bibnamefont
  {Gusakov}}, \bibinfo {author} {\bibfnamefont {E.~M.}\ \bibnamefont {Kantor}},
  \bibinfo {author} {\bibfnamefont {A.~I.}\ \bibnamefont {Chugunov}},\ and\
  \bibinfo {author} {\bibfnamefont {L.}~\bibnamefont {Gualtieri}},\ }\bibfield
  {title} {\bibinfo {title} {{Dissipation in relativistic superfluid neutron
  stars}},\ }\href {https://doi.org/10.1093/mnras/sts129} {\bibfield  {journal}
  {\bibinfo  {journal} {Mon. Not. Roy. Astron. Soc.}\ }\textbf {\bibinfo
  {volume} {428}},\ \bibinfo {pages} {1518} (\bibinfo {year} {2013})},\ \Eprint
  {https://arxiv.org/abs/1211.2452} {arXiv:1211.2452 [astro-ph.SR]}
  \BibitemShut {NoStop}%
\end{thebibliography}%

\end{document}